\documentclass[a4paper,aps,prd,10pt,preprintnumbers,showpacs,showkeys,twocolumn,superscriptaddress,nofootinbib,amsmath,amssymb]{revtex4-1}
\usepackage{cmap}
\usepackage[utf8]{inputenc}
\usepackage[T1]{fontenc}
\usepackage{graphicx,orcidlink,xcolor,ulem}
\usepackage{xcolor}

\begin{document}

\title{Bonanno-Reuter regular black hole: quasi-resonances, grey-body factors and absorption cross-sections of a massive scalar field}

\author{Zainab Malik}\email{zainabmalik8115@outlook.com}
\affiliation{Institute of Applied Sciences and Intelligent Systems, H-15, Pakistan}

\begin{abstract}
We study quasinormal modes of a massive scalar field in the background of the regular, quantum-corrected Bonanno–Reuter black hole, which arises from the renormalization group improvement of the Schwarzschild solution within the framework of asymptotically safe gravity. The analysis is performed in both the time and frequency domains. We find that increasing the mass of the field leads to a strong suppression of the damping rate, and extrapolation to larger masses indicates the emergence of arbitrarily long-lived oscillations, or quasi-resonances. In the time domain, the late-time decay follows an asymptotic behavior differs from the power-law tails of the classical Schwarzschild case. Furthermore, we compute the grey-body factors and absorption cross-sections for the massive scalar field and show that the grey-body factors decrease as the field mass increases, effectively shifting the emitted radiation spectrum toward higher frequencies. 
\end{abstract}

\maketitle

\section{Introduction}

The analysis of quasinormal modes (QNMs)~\cite{Kokkotas:1999bd,Berti:2009kk,Konoplya:2011qq,Bolokhov:2025uxz} has become a cornerstone of modern black-hole physics, linking theoretical models with observational data from gravitational-wave detectors~\cite{LIGOScientific:2016aoc,LIGOScientific:2017vwq,LIGOScientific:2020zkf,KAGRA:2013rdx}. These characteristic oscillations encode the fundamental response of compact objects to perturbations and thus serve as direct probes of the spacetime geometry and the underlying theory of gravity.  

While much of the existing literature has concentrated on massless perturbations, massive fields reveal a much richer phenomenology, introducing new dynamical scales and long-lived behavior~\cite{Ohashi:2004wr,Konoplya:2004wg}.. Quasinormal spectra of massive fields of different spins have been analyzed across a broad range of models~\cite{Skvortsova:2025cah,Zhidenko:2006rs,Konoplya:2017tvu,Burikham:2017gdm,Ohashi:2004wr,Aragon:2020teq,Gonzalez:2022upu,Zhang:2018jgj,Ponglertsakul:2020ufm,Konoplya:2004wg}.
These studies have demonstrated that the mass parameter of the field modifies both the oscillatory and damping characteristics of the spectrum in nontrivial ways. In higher-dimensional or braneworld setups, an effective mass can naturally appear due to Kaluza–Klein couplings between the bulk and brane~\cite{Seahra:2004fg}. In other contexts, massive gravitons predicted by modified gravity models could manifest in the ultra-long-wavelength regime probed by Pulsar Timing Array (PTA) observations~\cite{Konoplya:2023fmh,NANOGrav:2023hvm}, making the theoretical understanding of massive perturbations directly relevant for experiment.  

One of the most intriguing features of massive fields is the possible emergence of arbitrarily long-lived oscillations—known as {\it quasi-resonances}—at certain discrete mass values~\cite{Ohashi:2004wr,Konoplya:2004wg}. This effect is remarkably universal, appearing for various field spins~\cite{Fernandes:2021qvr,Percival:2020skc,Konoplya:2017tvu}, in different black-hole spacetimes~\cite{Zinhailo:2018ska,Churilova:2020bql,Konoplya:2013rxa,Zhidenko:2006rs,Bolokhov:2023bwm}, and even for non-black-hole compact configurations such as wormholes~\cite{Churilova:2019qph}. Nevertheless, the phenomenon is not generic—there exist examples where massive fields do not produce infinitely long-lived modes~\cite{Zinhailo:2024jzt}. Another essential feature of massive perturbations is the distinct structure of late-time tails: unlike massless fields that decay via a power law, massive fields exhibit oscillatory tails with a power-law envelope, a behavior found in various backgrounds~\cite{Koyama:2001ee,Rogatko:2007zz,Lutfuoglu:2025hwh,Gibbons:2008rs,Koyama:2000hj,Jing:2004zb,Moderski:2001tk,Gibbons:2008gg,Koyama:2001qw}. Even in the case of nominally massless fields, external influences such as magnetic fields or background matter distributions may induce an effective mass term~\cite{Konoplya:2008hj,Wu:2015fwa,Kokkotas:2010zd}, reinforcing the importance of studying such systems.  

Parallel to these developments, much effort has been devoted to understanding quantum and semiclassical corrections to black-hole geometries. Among the most compelling frameworks is the \textit{asymptotic safety program} \cite{Niedermaier:2006wt} of quantum gravity, which postulates that gravity becomes non-perturbatively renormalizable near a high-energy fixed point. Within this context, Bonanno and Reuter~\cite{Bonanno:2000ep} proposed a renormalization-group improvement of the Schwarzschild solution, in which Newton’s coupling $G$ becomes scale-dependent, $G \to G(k)$, with the running scale $k$ tied to the curvature or inverse radial distance. The resulting geometry,
\begin{equation}
f(r)=1-\frac{2MG(r)}{r},
\end{equation}
exhibits a de Sitter–like core and is free from the classical singularity, while asymptotic flatness is recovered at large radii. The corresponding “quantum-corrected” black hole thus provides a simple yet physically motivated model of semiclassical spacetime regularization, consistent with asymptotic safety.  

From the viewpoint of perturbation dynamics, the Bonanno–Reuter black hole offers a particularly interesting testing ground. It captures key qualitative features of quantum gravitational corrections—such as a softened central region and scale-dependent effective coupling—while retaining a tractable analytical form. These features can significantly alter the propagation of test fields, modify the effective potential barrier, and hence influence both the quasinormal spectra and grey-body factors. Furthermore, understanding how the mass of the field interacts with the quantum correction scale may help identify potential observational signatures in future high-precision gravitational-wave or electromagnetic measurements.  

The investigation of quasinormal modes of \textit{massless} test fields in various quantum-corrected backgrounds, including those in asymptotically safe gravity, has been already carried out in several studies~\cite{Bolokhov:2024bke,Lutfuoglu:2025pzi,Bolokhov:2023dxq,Macedo:2024dqb,Bolokhov:2025egl}.

However, the spectrum of \textit{massive} perturbations in the Bonanno–Reuter spacetime remains unexplored. Given the variety of phenomena introduced by a nonzero mass term—ranging from quasi-resonances to slowly decaying oscillatory tails—it is natural and timely to extend the analysis to this case. The present paper addresses this gap by computing the quasinormal modes of a massive scalar field in the background of the Bonanno–Reuter black hole, highlighting the qualitative and quantitative differences from the massless limit.  

Finally, we note that the literature on quasinormal modes of quantum-corrected black holes is vast. For completeness, we mention only a representative subset of recent works~\cite{Skvortsova:2024msa,Konoplya:2017lhs,Heidari:2023ssx,Malik:2025dxn,Abdalla:2005hu,Chen:2023wkq,Moreira:2023cxy,Konoplya:2020der,Bonanno:2025dry,Baruah:2023rhd,Fu:2023drp,Konoplya:2024lch,Skvortsova:2023zca,Malik:2024elk,Bolokhov:2025lnt,Anacleto:2019tdj,Anacleto:2021qoe}, referring the reader to references therein for further developments.

In addition to the study of quasinormal spectra, it is of great interest to analyze the scattering and emission properties of the Bonanno--Reuter black hole through its grey-body factors and absorption cross-sections. Grey-body factors describe the probability that radiation generated near the horizon is transmitted to an asymptotic observer, thus providing the essential link between the near-horizon dynamics and the observable Hawking flux. In the context of quantum-corrected geometries, these quantities are particularly informative, as they encode how the running of Newton’s constant and the softening of the curvature at small radii modify the propagation of test fields through the effective potential barrier. After all, we use the calculated grey-body factors to test the correspondence between the latter and quasinormal modes \cite{Konoplya:2024lir}. The absorption cross-section, in turn, measures the effective area of interaction between the black hole and incident radiation and serves as an important diagnostic of deviations from the classical Schwarzschild behavior. For massive fields, the situation becomes even more intricate: the mass introduces a new scale into the problem, altering both the structure of the effective potential and the transmission probabilities. 

In addition, in the context of gauge/gravity duality, grey-body factors acquire an additional significance: they govern the transmission of bulk perturbations through the black-hole horizon and thus determine the frequency-dependent conductivity of the dual holographic superconductor or plasma ~\cite{Horowitz:2008bn,Herzog:2009xv,Konoplya:2009hv}. In particular, the poles of the transmission coefficient are directly related to the electromagnetic response functions on the boundary, providing a bridge between black-hole scattering and linear transport phenomena in strongly coupled systems.

The paper is organized as follows. In Sec.~\ref{sec:theory} we briefly review the renormalization-group improvement of the Schwarzschild metric leading to the Bonanno--Reuter geometry and discuss its basic properties. In Sec.~\ref{sec:massive_scalar} we derive the master equation governing the dynamics of a massive scalar field in this background and specify the relevant boundary conditions for quasinormal modes. Section~\ref{sec:methods} describes the numerical and semi-analytical methods used for the calculation of the quasinormal frequencies, namely the high-order WKB approach with Padé approximants and the time-domain integration scheme. Sec. \ref{sec:QNMs} discusses the obtained quasinormal modes and evolution of perturbations in time domain. In Sec.~\ref{sec:gbf} we introduce the definitions of grey-body factors and absorption cross-sections and compute them using the WKB transmission formalism, including a discussion of their relation to the quasinormal spectrum. Finally, Sec.~\ref{sec:conclusions} summarizes the obtained results and outlines their implications for the dynamics of massive fields in quantum-corrected black-hole spacetimes.

\section{The Bonanno--Reuter Quantum-Corrected Black Hole}
\label{sec:theory}

The Bonanno--Reuter (BR) geometry represents one of the most studied realizations of
a quantum-corrected black hole emerging from the framework of
asymptotically safe gravity~\cite{Bonanno:2000ep}.  
Within this approach, the Einstein--Hilbert action is treated as a running effective theory whose
couplings evolve according to a renormalization group (RG) flow.
The essential idea is that the gravitational constant becomes scale dependent, 
so that Newton’s coupling \(G\) acquires an explicit dependence on the RG momentum scale \(k\),
which, in a stationary and spherically symmetric setting, can be related to the inverse of a
characteristic length scale of the system, \(k \simeq \xi / r\), with \(\xi\) being an order-unity parameter.

The improved description of the Schwarzschild geometry then follows from
replacing the constant \(G_0\) in the classical lapse function by a running coupling \(G(r)\),
whose explicit form is obtained by matching the RG flow to the one-loop effective
field-theory result for the Newtonian potential.  
A convenient and widely used parametrization is
\begin{equation}
\label{eq:Gr}
G(r) = 
\frac{G_0\, r^{3}}
{r^{3} + \tilde{\omega}\, G_0\, \bigl(r + \gamma\, G_0 M\bigr)}\,,
\end{equation}
where \(G_0\) denotes the infrared (low-energy) value of Newton’s constant,
\(M\) is the ADM mass parameter, 
\(\tilde{\omega} = 118/(15\pi)\) encodes the one-loop coefficient obtained from quantum-field-theoretic matching,
and \(\gamma>0\) is a dimensionless interpolation parameter that reflects
the freedom in choosing the RG cutoff identification.

Substituting~\eqref{eq:Gr} into the Schwarzschild lapse function yields the
RG-improved metric ~\cite{Bonanno:2000ep},

\begin{eqnarray}
\label{eq:metric}
&&ds^{2}
= -f(r)\, dt^{2}
+ \frac{dr^{2}}{f(r)}
+ r^{2}\, d\Omega^{2},
\\\nonumber&&
f(r)
= 1 - \frac{2 M\, G(r)}{r}
= 1 - \frac{2 G_0 M\, r^{2}}
{r^{3} + \tilde{\omega} G_0 \bigl(r + \gamma G_0 M\bigr)}.
\end{eqnarray}

This form smoothly interpolates between the classical Schwarzschild spacetime at large
distances and a regular, de Sitter–like geometry near the origin.

At large radial distances, the effective coupling tends to its classical value, \(G(r) \simeq G_0\),
and the lapse function expands as
\begin{equation}
f(r) \;\approx\;
1 - \frac{2 G_0 M}{r}
+ \frac{2 \tilde{\omega} G_0^{2} M}{r^{3}}
+ \mathcal{O}(r^{-4}),
\end{equation}
revealing quantum-gravity corrections that scale as \(1/r^{3}\).
These subleading terms are numerically tiny for astrophysical black holes but become
relevant at microscopic scales.

In the opposite regime \(r \to 0\),
the function \(G(r)\) vanishes quadratically with \(r\),
\begin{equation}
G(r) \;\simeq\; 
\frac{r^{3}}{\tilde{\omega}\,(r + \gamma G_0 M)}
\;\approx\; \frac{r^{3}}{\tilde{\omega} \gamma G_0 M},
\end{equation}
leading to a de Sitter core with an effective cosmological constant.
This behavior ensures regularity of all curvature invariants at the center.

Depending on the mass \(M\) and the parameter \(\gamma\),
the lapse function \(f(r)\) may exhibit two, one, or no real positive roots,
corresponding respectively to a regular black hole with outer and inner horizons,
an extremal configuration, or a horizonless, particle-like remnant.
A critical mass \(M_{\mathrm{cr}}\) separates the black-hole and remnant regimes.
For \(M > M_{\mathrm{cr}}\), the geometry behaves as a standard asymptotically flat black hole;
for \(M \le M_{\mathrm{cr}}\), the object is completely regular and non-singular.

Throughout this paper we fix \(\tilde{\omega} = 118/(15\pi)\)
and adopt geometrized units \(G_0 = c = 1\).
The parameter \(\gamma\) is kept as a free quantity that controls the strength of
RG improvement and thereby the degree of quantum modification.
The resulting spacetime serves as an effective, regularized model for studying
how the asymptotic-safety framework modifies wave propagation and
observable quantities such as grey-body factors and absorption cross-sections.

\section{Massive scalar-field perturbations}
\label{sec:massive_scalar}

We now consider the dynamics of a minimally coupled scalar field of rest mass $\mu$
propagating in the quantum-corrected Bonanno--Reuter spacetime.
The field evolution is governed by the Klein--Gordon equation,
\begin{equation}
\Box \Phi - \mu^{2} \Phi = 0,
\qquad
\Box \equiv g^{\mu\nu} \nabla_{\mu} \nabla_{\nu},
\label{eq:KG}
\end{equation}
where $\nabla_{\mu}$ denotes the covariant derivative compatible with the metric~\eqref{eq:metric}.
The term $\mu^{2}\Phi$ introduces a finite reduced Compton wavelength
$\lambdabar_{c}=1/\mu$, which adds a new characteristic scale to the problem.

To exploit the spherical symmetry of the background, we perform a separation of variables using
spherical harmonics $Y_{\ell m}(\theta,\phi)$, which satisfy
\begin{eqnarray}
&&\nabla^{2}_{\!_{S^{2}}} Y_{\ell m}(\theta,\phi)
= - \ell(\ell+1)\, Y_{\ell m}(\theta,\phi),
\\\nonumber
&&\ell = 0,1,2,\ldots,\qquad m=-\ell,\ldots,+\ell.
\end{eqnarray}
The field can then be decomposed as
\begin{equation}
\Phi(t,r,\theta,\phi)
= \frac{1}{r}\,\Psi_{\ell}(r)\, Y_{\ell m}(\theta,\phi)\, e^{- i \omega t},
\label{eq:ansatz}
\end{equation}
where the radial function $\Psi_{\ell}(r)$ encodes the essential dynamics.
Substituting~\eqref{eq:ansatz} into~\eqref{eq:KG} and using the line element~\eqref{eq:metric}
one obtains, after standard manipulations, the one-dimensional wave equation
\begin{equation}
\frac{d^{2}\Psi_{\ell}}{dr_{*}^{2}}
+ \Bigl[\omega^{2}-V_{\ell}(r)\Bigr] \Psi_{\ell} = 0,
\label{eq:RadialEq}
\end{equation}
where the tortoise coordinate $r_{*}$ is defined through
\begin{equation}
\frac{dr_{*}}{dr} = \frac{1}{f(r)},
\qquad
r_{*} \;\to\;
\begin{cases}
-\infty, & r \to r_{+},\\[2pt]
+\infty, & r \to \infty,
\end{cases}
\end{equation}
with $r_{+}$ denoting the event-horizon radius.

The effective potential $V_{\ell}(r)$ takes the form
\begin{equation}
V_{\ell}(r)
= f(r)
\!\left[
\mu^{2}
+ \frac{\ell(\ell+1)}{r^{2}}
+ \frac{f'(r)}{r}
\right],
\label{eq:Vscalar}
\end{equation}
where the prime indicates differentiation with respect to $r$.
The first two terms represent, respectively, the mass contribution and the centrifugal barrier,
while the last term originates from the background curvature and vanishes in flat spacetime.
For $\mu = 0$, this potential reduces to the standard form known from the Schwarzschild case,
whereas for $\mu \ne 0$ an additional long-range contribution appears that
significantly modifies both the quasinormal spectrum and the late-time behavior.

The physical boundary conditions depend on the asymptotic structure of the spacetime.
In the asymptotically flat case, the correct quasinormal-mode (QNM) conditions require
purely ingoing waves at the event horizon and purely outgoing waves at spatial infinity:
\begin{equation}
\Psi_{\ell}(r)
\;\propto\;
\begin{cases}
e^{- i \omega r_{*}}, & r \to r_{+}\quad (r_{*}\to -\infty),\\[6pt]
e^{+ i \Omega r_{*}}, & r \to \infty\quad (r_{*}\to +\infty),
\end{cases}
\label{eq:BCs}
\end{equation}
where the wave number
\begin{equation}
\Omega = \sqrt{\omega^{2} - \mu^{2}},
\label{eq:Omega}
\end{equation}
is defined such that $\mathrm{Re}(\Omega)$ and $\mathrm{Re}(\omega)$ have the same sign~\cite{Zhidenko:2006rs}.

The discrete complex frequencies $\omega = \omega_{\mathrm{R}} - i\,\omega_{\mathrm{I}}$
satisfying~\eqref{eq:RadialEq} and~\eqref{eq:BCs}
represent the quasinormal spectrum of the quantum-corrected black hole.
The real part $\omega_{\mathrm{R}}$ determines the oscillation frequency of the perturbation,
while the imaginary part $\omega_{\mathrm{I}}>0$ describes the damping rate.
In certain parameter ranges, typically at specific combinations of $(\mu,\ell)$,
$\omega_{\mathrm{I}}$ can approach zero, giving rise to extremely long-lived oscillations
known as {\it quasi-resonances}.
At late times, instead of the standard power-law tail characteristic of massless fields,
the signal exhibits an oscillatory decay modulated by a power-law envelope,
reflecting the dispersive nature of massive wave propagation in curved spacetime.

\begin{figure}
\resizebox{\linewidth}{!}{\includegraphics{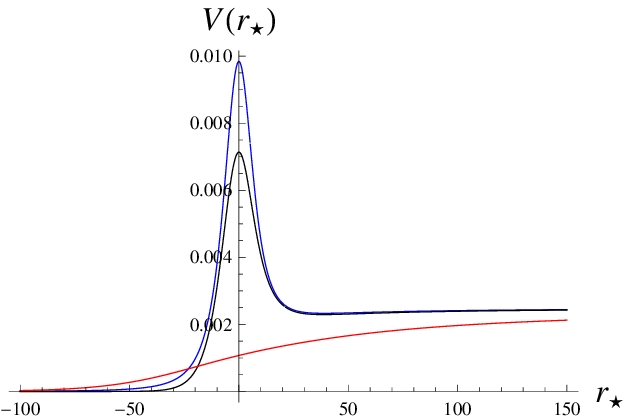}}
\caption{Effective potentials for $\ell=0$, $\gamma=0.1$, $\mu=0.05$: $M=1.66$ (blue, top), $M=2$ (black,middle), $M=10$ (red,bottom).}\label{fig:Potential1}
\end{figure}

\begin{figure}
\resizebox{\linewidth}{!}{\includegraphics{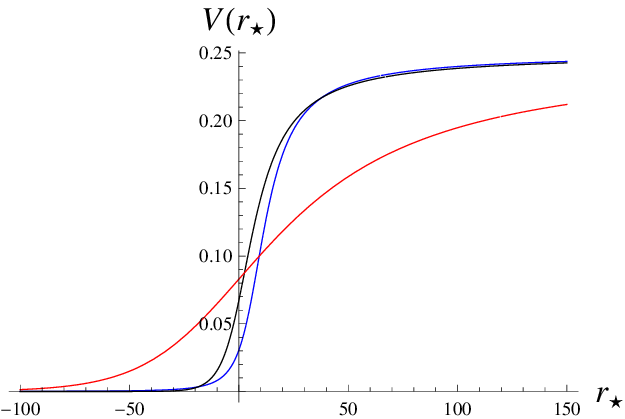}}
\caption{Effective potentials for $\ell=0$, $\gamma=0.1$, $\mu=0.5$: $M=1.66$ (blue, top), $M=2$ (black,middle), $M=10$ (red,bottom).}\label{fig:Potential2}
\end{figure}

\vspace{5mm}
\section{Methods}\label{sec:methods} 
\subsection{WKB method and Padé resummation}

A powerful semi-analytic approach to the computation of quasinormal modes of black holes
is based on the Wentzel–Kramers–Brillouin (WKB) approximation.
Originally formulated for potential scattering problems,
this technique was adapted to black-hole perturbations in the seminal works
of Schutz and Will~\cite{Schutz:1985km} and later improved by Iyer and Will~\cite{Iyer:1986np}.
Higher-order generalizations and Padé resummations
have since rendered the method highly accurate for a wide range of effective potentials
\cite{Konoplya:2003ii,Konoplya:2019hlu,Matyjasek:2017psv}.

The method exploits the fact that the effective potential $V(r)$ possesses a single
well-defined peak between the event horizon and spatial infinity.
In the neighborhood of this maximum, denoted by $r=r_0$,
the potential can be expanded in a Taylor series with respect to the tortoise 
coordinate $r_{*}$:
\begin{eqnarray}
&&\!\!\!\!V(r_*) = V_0 + \frac{V_0^{(2)}}{2}(r_* - r_{*0})^2
+ \frac{V_0^{(3)}}{6}(r_* - r_{*0})^3 + \cdots ,
\nonumber\\
&&V_0^{(n)} \equiv \left. \frac{d^{n}V}{dr_{*}^{n}}\right|_{r_{*0}}.
\end{eqnarray}
Matching the WKB solutions across the classical turning points
and requiring purely outgoing behavior at both boundaries
leads to a quantization condition for the complex frequencies:
\begin{equation}
\frac{i(\omega^{2}-V_0)}{\sqrt{-2V_0^{(2)}}}
- \sum_{k=2}^{N} \Lambda_k(\{V_0^{(j)}\},n)
= n + \frac{1}{2}, \quad n = 0,1,2,\ldots,
\label{eq:WKBquantization}
\end{equation}
where $n$ is the overtone number, and the correction terms $\Lambda_k$
are known algebraic expressions involving the higher derivatives of the potential.
Explicit forms of $\Lambda_k$ up to the 13th order can be found in
Refs.~\cite{Iyer:1986np,Konoplya:2003ii,Matyjasek:2017psv}.

Equation~\eqref{eq:WKBquantization} can be viewed as an implicit polynomial relation for $\omega^{2}$.
Because the WKB expansion is asymptotic, truncating it at finite order $N$
does not necessarily improve accuracy monotonically with increasing $N$.
A robust way to improve convergence is to treat the WKB series as a formal
power expansion in a bookkeeping parameter $\epsilon$ and apply a Padé resummation.
Defining
\begin{equation}
\omega^{2}(\epsilon)
= V_0 - i\sqrt{-2V_0^{(2)}}\left(\epsilon\!\left(n+\tfrac{1}{2}\right)
+ \sum_{k=2}^{N} \epsilon^{k}\Lambda_k\right),
\label{eq:omegaseries}
\end{equation}
one constructs the Padé rational approximation
\begin{equation}
\mathcal{P}_{\tilde{m}/\tilde{n}}(\epsilon)
= \frac{\displaystyle \sum_{j=0}^{\tilde{m}} a_j \epsilon^{j}}
{\displaystyle  \sum_{k=1}^{\tilde{n}} b_k \epsilon^{k}},
\qquad
\tilde{m} + \tilde{n} = N,
\label{eq:pade}
\end{equation}
whose Taylor expansion matches~\eqref{eq:omegaseries} up to order $\mathcal{O}(\epsilon^{N+1})$.
The final Padé estimate for the quasinormal frequency is then
\begin{equation}
\omega = \sqrt{\,\mathcal{P}_{\tilde{m}/\tilde{n}}(1)\,}.
\label{eq:padeomega}
\end{equation}

Balanced approximants, such as $[\tilde{m}/\tilde{n}] = [3/3]$ or $[4/4]$
for $N=6$ or $N=8$ respectively, typically yield the most stable results.
The residual difference between several near-balanced approximants
provides a reliable internal estimate of the order of the numerical uncertainty.
When compared with exact results obtained by the Frobenius (Leaver) method,
the sixth- or seventh-order WKB–Padé approach usually reproduces
the real and imaginary parts of the dominant mode with relative errors
below $0.1\%$ for $\ell \ge 1$, while maintaining
good qualitative accuracy even for the fundamental monopole mode.

It is important to note that the WKB approximation is valid only
for potentials possessing two well-defined turning points,
and it assumes that the wave oscillates rapidly within the barrier region.
Therefore, it performs best for low overtones ($n < \ell$)
and for relatively small field masses $\mu$,
when the effective potential preserves a single peak.
In the regime of large $\mu$, where the potential does not have a single peak barrier form, one must rely on more precise
techniques such as the Frobenius \cite{Leaver:1985ax} or time-domain integration methods. The WKB method has been employed in numerous studies, and comparisons with results obtained by more accurate approaches generally demonstrate good agreement for high multipole numbers, However, for higher overtones and certain ranges of parameters, the method may sometimes exhibit insufficient accuracy (see \cite{Batic:2024zbx,Batic:2024vsb,Batic:2024vwm,Batic:2025hgp} for examples of both types of behavior).

Throughout this work we employ the sixth- and seventh-order
WKB formalism together with Padé approximants $[3/3]$ and $[3/4]$,
which have been shown in many contexts to provide
an optimal accuracy.

\subsection{Time-domain integration method}

A complementary approach to the WKB formalism is provided by the direct numerical
integration of the perturbation equation in the time domain.
This method allows one to study the full dynamical evolution of the field,
including the transition from the initial transient to the quasinormal ringing
and the subsequent late-time tail.
It is particularly useful in regimes where semi-analytic methods such as the WKB
approximation lose accuracy, for example for low multipoles, 
or massive fields whose effective potential is not of the standard barrier form.

Starting from the wave-like equation
\begin{equation}
\frac{\partial^2 \Psi}{\partial t^2}
- \frac{\partial^2 \Psi}{\partial r_*^2}
+ V(r)\, \Psi = 0,
\label{eq:waveTD}
\end{equation}
it is convenient to adopt the characteristic (light-cone) coordinates
\begin{equation}
u = t - r_*, \qquad v = t + r_*,
\end{equation}
in which the equation takes the form
\begin{equation}
4\,\frac{\partial^2 \Psi}{\partial u \partial v}
+ V(r)\, \Psi = 0.
\label{eq:waveUV}
\end{equation}
The radial coordinate $r$ at each grid point is recovered implicitly
from the relation between $r$ and $r_*$,
\begin{equation}
r_*(r) = \int^r \frac{dr'}{f(r')}.
\end{equation}

Discretization on a uniform $(u,v)$ grid leads to a simple and stable finite-difference scheme
originally proposed by Gundlach, Price, and Pullin~\cite{Gundlach:1993tp}.
Labeling the grid points by $N=(u+\Delta, v+\Delta)$,
$E=(u+\Delta,v)$, $W=(u,v+\Delta)$, and $S=(u,v)$,
the evolution equation is approximated as
\begin{equation}
\Psi_N = \Psi_E + \Psi_W - \Psi_S
- \frac{\Delta^2}{8}\,V(r_S)\,\bigl(\Psi_E + \Psi_W\bigr)
+ \mathcal{O}(\Delta^4),
\label{eq:FDscheme}
\end{equation}
where $\Delta$ is the grid step.
Initial data are specified on the two null surfaces $u=u_0$ and $v=v_0$.
In most cases, one chooses a Gaussian pulse of compact support as initial perturbation.

The field is then evolved numerically along the grid using~\eqref{eq:FDscheme},
and the signal $\Psi(t,r)$ is recorded at a fixed spatial position outside the potential barrier.
The obtained waveform contains three distinct regimes:
an early transient depending on the initial pulse,
an exponentially damped oscillatory stage corresponding to quasinormal ringing,
and a late-time tail governed by backscattering from distant regions of the effective potential.

To determine the quasinormal frequencies from the time-domain data,
one typically fits the intermediate portion of the waveform
(where the quasinormal ringing dominates)
to a sum of damped exponentials of the form
\begin{equation}
\Psi(t) = \sum_{j=1}^{N} A_j \, e^{-i\omega_j t},
\qquad
\omega_j = \omega_{R,j} - i \omega_{I,j},
\label{eq:PronyFit}
\end{equation}
where $\omega_{R,j}$ and $\omega_{I,j}$ are the real oscillation frequency
and damping rate of each mode, respectively.
This fitting can be performed efficiently using the Prony or matrix-pencil method,
which reconstructs the complex frequencies $\omega_j$
directly from discrete time-series data without requiring a Fourier transform
(see, e.g., 
\cite{Lutfuoglu:2025qkt,Varghese:2011ku,Churilova:2019qph,Dubinsky:2024jqi,Malik:2024bmp,Qian:2022kaq,Konoplya:2020jgt,Bolokhov:2023ruj,Cuyubamba:2016cug,Lutfuoglu:2025bsf,Malik:2025ava,Ishihara:2008re,Konoplya:2013sba,Stuchlik:2025mjj,Bolokhov:2026eqf,Bolokhov:2025fto}).

For massless fields, the quasinormal ringing is followed by a power-law tail,
whereas for massive perturbations the tail becomes oscillatory
with a slowly decaying envelope, reflecting the finite Compton wavelength of the field.
In such cases, the time-domain method captures both the transition
from the quasinormal stage to the oscillatory tail and
provides an important cross-check of the frequency-domain calculations.

The time-domain approach offers several advantages:
it is numerically stable, does not require an explicit form of the wavefunction at infinity,
and remains applicable even when the effective potential has multiple peaks
or does not decay monotonically.
However, its precision for extracting higher overtones
is limited by the exponentially decreasing amplitude of subdominant modes.
For the fundamental mode,
the frequencies obtained from the time-domain integration
typically agree with WKB or Frobenius results within a small fraction of a percent,
thus providing a reliable verification of semi-analytic methods.

\section{Quasinormal modes}\label{sec:QNMs}

\begin{figure*}
\resizebox{\linewidth}{!}{
\includegraphics{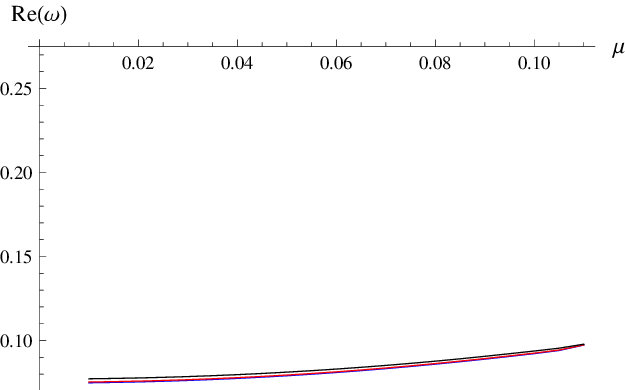}
\includegraphics{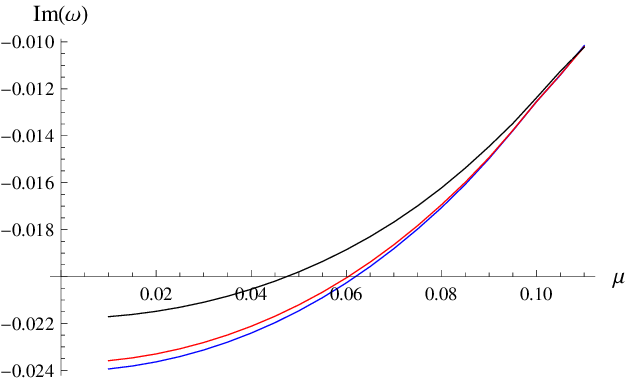}
}
\caption{Real and imaginary part of the dominant ($n=0$) quasinormal modes of the $\ell=1$, $M=4$  case  for the Bonanno-Reuter black hole calculated by the 6th-order WKB formula with Padé resummation ($[\tilde{m}/\tilde{n}] = [3/3]$) as functions of $\mu$, $\gamma=0.01$ (blue), $\gamma=1$ (red), $\gamma=9/2$ (black).}\label{fig:qnms1}
\end{figure*}

\begin{figure*}
\resizebox{\linewidth}{!}{
\includegraphics{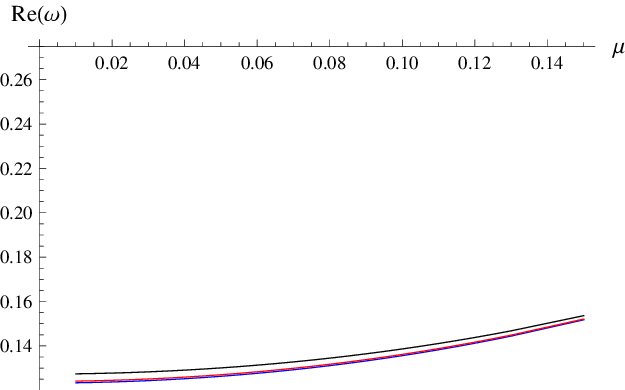}
\includegraphics{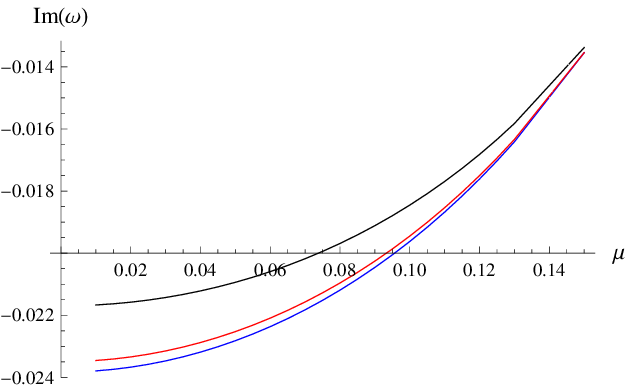}
}
\caption{Real and imaginary part of the dominant ($n=0$) quasinormal modes of the $\ell=2$, $M=4$ case for the Bonanno-Reuter black hole calculated by the 6th-order WKB formula with Padé resummation ($[\tilde{m}/\tilde{n}] = [3/3]$) as functions of $\mu$, $\gamma=0.01$ (blue), $\gamma=1$ (red), $\gamma=9/2$ (black).}\label{fig:qnms2}
\end{figure*}

\begin{figure*}
\resizebox{\linewidth}{!}{
\includegraphics{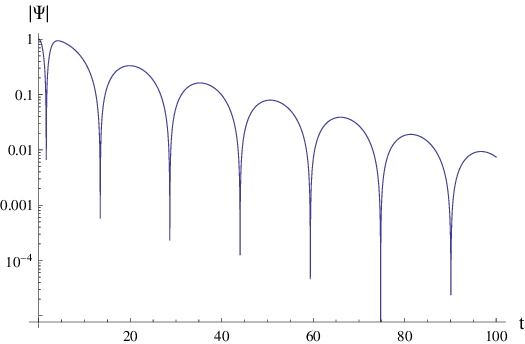}
\includegraphics{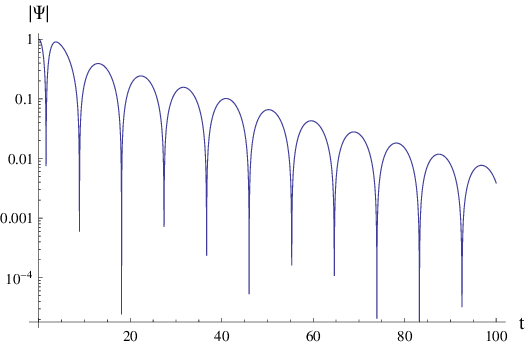}
}
\caption{Semi-logarithmic time-domain profiles for $\ell=1$ perturbations (left) and $\ell=2$ ones (right). Here $M=1.66$, $\gamma=0.1$. For the left plot WKB gives $\omega=0.203056- 0.046915 i$, while the Prony method gives $\omega =0.20446 - 0.04648 i$ making  the relative error less than one percent. For the right plot we have $\omega = 0.3376 - 0.0462 i$ by the Prony method and $\omega =0.3376 - 0.0461 i$ by the WKB method.}\label{fig:TD}
\end{figure*}

\begin{figure}
\resizebox{\linewidth}{!}{\includegraphics{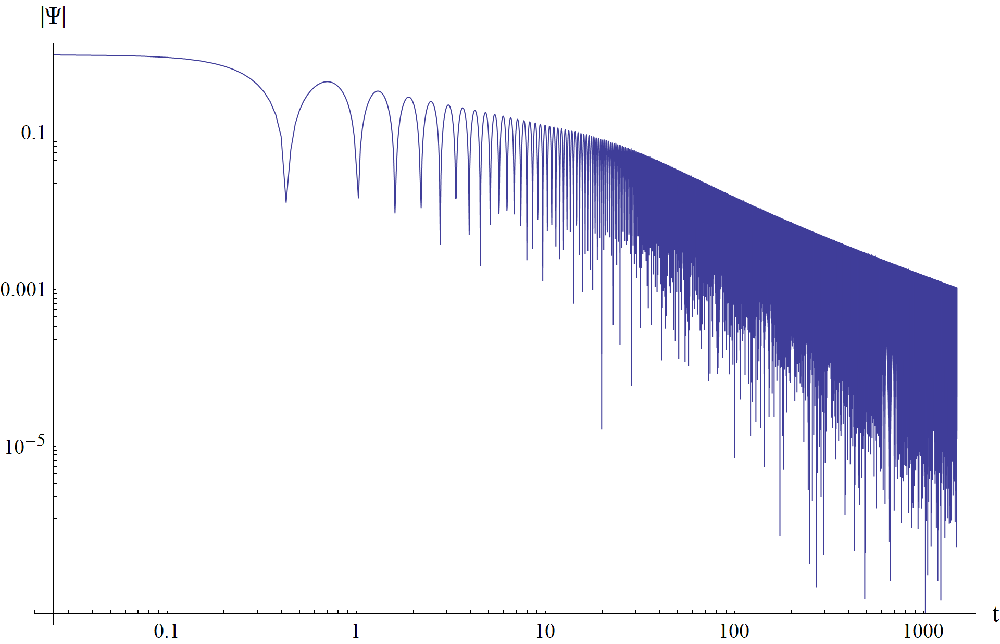}}
\caption{Logarithmic time-domain profiles for $\ell=0$ perturbations, $M=1.66$, $\gamma=0.1$, $\mu=10$. The asymptotic tail is close to $\propto t^{-7/8}$.}\label{fig:TD2}
\end{figure}

\begin{figure}
\resizebox{\linewidth}{!}{\includegraphics{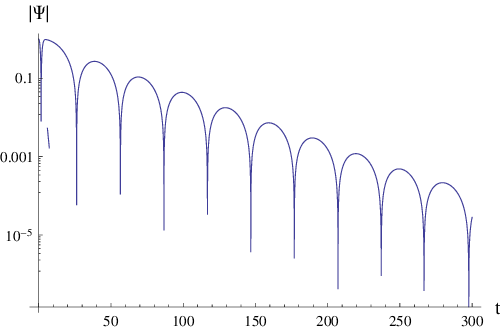}}
\caption{Logarithmic time-domain profiles for $\ell=1$ perturbations with $M=3$, $\gamma=1$, and $\mu=0.05$. Applying the Prony method to the early ringdown stage yields $\omega = 0.1055 - 0.0288 i$, which is in very good agreement with the WKB result $\omega =0.105578 - 0.028895 i$. However, fitting the signal at later times, around $t\sim 200$--$300$, already appears to be contaminated by intermediate tails.} \label{fig:TD3}
\end{figure}

\begin{table}
\begin{tabular}{c l c c l}
\hline
\hline
$\ell$ & $M$  & WKB-6, $m=3$ & WKB-7, $m=3$ &  $\delta (\%)$ \\
\hline
$0$ & $1.66$ & $0.072135-0.051121 i$ & $0.072446-0.050962 i$ & $0.395\%$\\
$0$ & $1.8$ & $0.067554-0.048366 i$ & $0.067691-0.048184 i$ & $0.274\%$\\
$0$ & $2.$ & $0.060810-0.045860 i$ & $0.061476-0.046157 i$ & $0.957\%$\\
$0$ & $2.5$ & $0.047705-0.039162 i$ & $0.047467-0.039507 i$ & $0.679\%$\\
$0$ & $3.$ & $0.038933-0.033564 i$ & $0.038933-0.033559 i$ & $0.010\%$\\
$0$ & $3.5$ & $0.032980-0.029105 i$ & $0.032951-0.029046 i$ & $0.150\%$\\
$0$ & $4.$ & $0.028653-0.025639 i$ & $0.028629-0.025585 i$ & $0.154\%$\\
$0$ & $5.$ & $0.022734-0.020663 i$ & $0.022722-0.020624 i$ & $0.134\%$\\
$0$ & $10.$ & $0.011239-0.010427 i$ & $0.011237-0.010412 i$ & $0.101\%$\\
$1$ & $1.66$ & $0.203056-0.046915 i$ & $0.203033-0.046907 i$ & $0.0117\%$\\
$1$ & $1.8$ & $0.182658-0.046637 i$ & $0.182642-0.046638 i$ & $0.0083\%$\\
$1$ & $2.$ & $0.160197-0.044213 i$ & $0.160190-0.044215 i$ & $0.0047\%$\\
$1$ & $2.5$ & $0.123658-0.037183 i$ & $0.123658-0.037183 i$ & $0.0003\%$\\
$1$ & $3.$ & $0.101250-0.031574 i$ & $0.101251-0.031573 i$ & $0.0010\%$\\
$1$ & $3.5$ & $0.085914-0.027321 i$ & $0.085915-0.027321 i$ & $0.0015\%$\\
$1$ & $4.$ & $0.074698-0.024039 i$ & $0.074699-0.024039 i$ & $0.0016\%$\\
$1$ & $5.$ & $0.059324-0.019347 i$ & $0.059325-0.019347 i$ & $0.0016\%$\\
$1$ & $10.$ & $0.029383-0.009744 i$ & $0.029384-0.009744 i$ & $0.001\%$\\
\hline
\hline
\end{tabular}
\caption{Quasinormal modes ($n=0$) of the scalar  perturbations of the Bonanno-Reuter black hole $\gamma=0.1$, $\mu=0$ calculated using the WKB formula at different orders together with the difference between them.}
\end{table}

\begin{table}
\begin{tabular}{c l c c l}
\hline
\hline
$\ell$ & $M$  & WKB-6, $m=3$ & WKB-7, $m=3$ &  $\delta (\%)$ \\
\hline
$0$ & $2.19$ & $0.055329-0.035954 i$ & $0.055430-0.035824 i$ & $0.250\%$\\
$0$ & $2.25$ & $0.054154-0.034991 i$ & $0.054189-0.034911 i$ & $0.135\%$\\
$0$ & $2.5$ & $0.049781-0.032022 i$ & $0.049775-0.032122 i$ & $0.169\%$\\
$1$ & $2.19$ & $0.149324-0.035982 i$ & $0.149323-0.035982 i$ & $0.0003\%$\\
$1$ & $2.25$ & $0.144733-0.035680 i$ & $0.144734-0.035680 i$ & $0.0001\%$\\
$1$ & $2.5$ & $0.128427-0.033682 i$ & $0.128433-0.033681 i$ & $0.0047\%$\\
$1$ & $3.$ & $0.105578-0.028895 i$ & $0.105583-0.028895 i$ & $0.0046\%$\\
$1$ & $3.5$ & $0.090354-0.024684 i$ & $0.090357-0.024684 i$ & $0.0029\%$\\
$1$ & $4.$ & $0.079480-0.021208 i$ & $0.079480-0.021208 i$ & $0.0007\%$\\
$1$ & $4.5$ & $0.071344-0.018322 i$ & $0.071342-0.018322 i$ & $0.0021\%$\\
$1$ & $5.$ & $0.065056-0.015882 i$ & $0.065051-0.015882 i$ & $0.0064\%$\\
\hline
\hline
\end{tabular}
\caption{Quasinormal modes ($n=0$) of the scalar  perturbations the Bonanno-Reuter black hole ($\gamma=1$, $\mu=0.05$) calculated using the WKB formula at different orders together with the difference between them.}
\end{table}

\begin{table}
\begin{tabular}{c l c c l}
\hline
\hline
$\ell$ & $M$  & WKB-6, $m=3$ & WKB-7, $m=3$ &  $\delta (\%)$ \\
\hline
$2$ & $3.5$ & $0.145471-0.025736 i$ & $0.145472-0.025735 i$ & $0.0004\%$\\
$2$ & $3.6$ & $0.141312-0.025039 i$ & $0.141312-0.025038 i$ & $0.0004\%$\\
$2$ & $3.7$ & $0.137406-0.024370 i$ & $0.137406-0.024370 i$ & $0.0003\%$\\
$2$ & $3.8$ & $0.133730-0.023730 i$ & $0.133731-0.023729 i$ & $0.0003\%$\\
$2$ & $3.9$ & $0.130266-0.023115 i$ & $0.130266-0.023115 i$ & $0.0003\%$\\
$2$ & $4.$ & $0.126993-0.022525 i$ & $0.126994-0.022525 i$ & $0.0002\%$\\
$2$ & $4.5$ & $0.113028-0.019907 i$ & $0.113028-0.019907 i$ & $0.0001\%$\\
$2$ & $5.$ & $0.102104-0.017740 i$ & $0.102104-0.017740 i$ & $0\%$\\
$2$ & $10.$ & $0.056614-0.006768 i$ & $0.056613-0.006767 i$ & $0.0024\%$\\
$3$ & $3.5$ & $0.201403-0.026068 i$ & $0.201403-0.026068 i$ & $0\%$\\
$3$ & $3.6$ & $0.195543-0.025386 i$ & $0.195543-0.025386 i$ & $0\%$\\
$3$ & $3.7$ & $0.190036-0.024734 i$ & $0.190036-0.024734 i$ & $0\%$\\
$3$ & $3.8$ & $0.184851-0.024109 i$ & $0.184851-0.024109 i$ & $0\%$\\
$3$ & $3.9$ & $0.179960-0.023510 i$ & $0.179960-0.023510 i$ & $0\%$\\
$3$ & $4.$ & $0.175338-0.022936 i$ & $0.175338-0.022936 i$ & $0\%$\\
$3$ & $4.5$ & $0.155572-0.020396 i$ & $0.155572-0.020396 i$ & $0\%$\\
$3$ & $5.$ & $0.140051-0.018307 i$ & $0.140051-0.018307 i$ & $0\%$\\
$3$ & $10.$ & $0.073733-0.008164 i$ & $0.073733-0.008164 i$ & $0\%$\\
\hline
\hline
\end{tabular}
\caption{Quasinormal modes ($n=0$) of the scalar  perturbations the Bonanno-Reuter black hole $\gamma=9/2$, $\mu=0.05$ calculated using the WKB formula at different orders together with the difference between them.}
\end{table}

\begin{table}
\begin{tabular}{c l c c l}
\hline
\hline
$\ell$ & $M$  & WKB-6, $m=3$ & WKB-7, $m=3$ &  $\delta (\%)$ \\
\hline
$2$ & $3.5$ & $0.143810-0.073456 i$ & $0.143794-0.073448 i$ & $0.0115\%$\\
$2$ & $3.6$ & $0.139800-0.071177 i$ & $0.139789-0.071163 i$ & $0.0110\%$\\
$2$ & $3.7$ & $0.136033-0.068981 i$ & $0.136032-0.068966 i$ & $0.0095\%$\\
$2$ & $3.8$ & $0.132493-0.066856 i$ & $0.132497-0.066851 i$ & $0.0045\%$\\
$2$ & $3.9$ & $0.129173-0.064811 i$ & $0.129167-0.064812 i$ & $0.0045\%$\\
$2$ & $4.$ & $0.126043-0.062852 i$ & $0.126024-0.062844 i$ & $0.0142\%$\\
$2$ & $4.5$ & $0.112664-0.054049 i$ & $0.112664-0.053963 i$ & $0.0692\%$\\
$2$ & $5.$ & $0.102211-0.046566 i$ & $0.102273-0.046421 i$ & $0.141\%$\\
$3$ & $3.5$ & $0.201687-0.075668 i$ & $0.201688-0.075667 i$ & $0.0007\%$\\
$3$ & $3.6$ & $0.196009-0.073493 i$ & $0.196010-0.073492 i$ & $0.0007\%$\\
$3$ & $3.7$ & $0.190683-0.071404 i$ & $0.190684-0.071403 i$ & $0.0007\%$\\
$3$ & $3.8$ & $0.185677-0.069398 i$ & $0.185678-0.069397 i$ & $0.0007\%$\\
$3$ & $3.9$ & $0.180963-0.067470 i$ & $0.180964-0.067469 i$ & $0.0007\%$\\
$3$ & $4.$ & $0.176517-0.065617 i$ & $0.176517-0.065616 i$ & $0.0008\%$\\
$3$ & $4.5$ & $0.157600-0.057341 i$ & $0.157601-0.057342 i$ & $0.0005\%$\\
$3$ & $5.$ & $0.142885-0.050417 i$ & $0.142884-0.050415 i$ & $0.0011\%$\\
\hline
\hline
\end{tabular}
\caption{Quasinormal modes ($n=1$) of the scalar perturbations the Bonanno-Reuter black hole $\gamma=9/2$, $\mu=0.1$ calculated using the WKB formula at different orders together with the difference between them.}
\end{table}

As mentioned before, in the present work we employ the higher-order WKB formalism combined with Padé resummation, which significantly enhances convergence and extends the range of validity of the approximation.  
Specifically, we use expnasion of 6th and 7th orders  with balanced Padé approximants such as $[3/3]$ or $[3/4]$, following the well-tested prescriptions developed in a series of works~\cite{Dubinsky:2024hmn,Skvortsova:2023zmj,Konoplya:2020fwg,Bronnikov:2021liv,Malik:2023bxc,Skvortsova:2024atk,Bolokhov:2023bwm,Dubinsky:2025bvf,Konoplya:2021ube,Dubinsky:2024mwd,Dubinsky:2024nzo,Skvortsova:2024wly,Dubinsky:2024rvf,Lutfuoglu:2025hjy,Malik:2024nhy,Churilova:2021tgn,Konoplya:2019xmn,Bronnikov:2019sbx,Bolokhov:2024ixe,Dubinsky:2025fwv}.  

The introduction of a nonzero field mass $\mu$ modifies the asymptotic structure of the effective potential, which may lose its canonical single-barrier shape.  
As $\mu$ increases, the potential ceases to have precisely two classical turning points, and at some large $\mu$ does not have a maximum, so that the assumptions underlying the WKB formalism break down, rendering the method unreliable.  Consequently, the applicability of the WKB–Padé scheme must be restricted to the parameter domain in which the potential exhibits a single well-defined maximum.  
In this regime, the method yields reasonable estimations for both the real oscillation frequency and the damping rate, while also allowing for efficient cross-validation against numerical techniques such as time-domain integration.  
For large field masses, by contrast, the time-domain evolution remains the only reliable approach, as it makes no assumptions about the shape of the potential and automatically captures possible transitions from oscillatory to quasi-stationary behavior.

Although the WKB method guarantees only asymptotic convergence, rather than strict convergence at each order, the presence of a plateau across several successive orders typically provides a reliable estimate of the relative error. From Tables~I–IV we can compare the results obtained at different WKB orders and observe that, as will be shown later, the deviation from the time-domain integration data is of the same magnitude.

From Figs.~\ref{fig:qnms1} and~\ref{fig:qnms2} we observe that as the field mass $\mu$ increases, the damping rate of oscillations decreases significantly. In the regime of larger $\mu$, the damping rates corresponding to different values of the quantum parameter tend to converge toward the Schwarzschild limit. For the latter case, it is well established that arbitrarily long-lived oscillations—quasi-resonances—appear at certain threshold values of the field mass. Hence, it is reasonable to conclude that a similar phenomenon occurs for the Bonanno–Reuter black hole. Nevertheless, within the WKB framework this regime cannot be fully captured, since the effective potential loses its single-peak structure when approaching the quasi-resonant limit.  

The comparison between the WKB data and the results of time-domain integration demonstrates excellent agreement: the relative deviation between the two methods remains below one percent (see fig. \ref{fig:TD}).

At  very late times, when the evolution enters the regime $\mu t \gg 1/(\mu^{2} M^{2})$, the character of the massive-field dynamics changes qualitatively.  
In this asymptotic domain, the familiar exponential or power-law decay is replaced by a slowly oscillating inverse-power behavior—commonly known as the  asymptotic tail—which represents the long-range interference between backscattered massive modes~\cite{Koyama:2001qw,Burko:2004jn,Koyama:2001ee,Lutfuoglu:2025qkt}.  
This stage dominates the signal once the quasinormal ringing has completely faded, reflecting the persistence of massive excitations in the far zone of the black-hole spacetime.

For the Schwarzschild, Reissner-Nordström and some other spacetimes, the asymptotic decay law is:
\begin{equation}\label{asymptotictail}
\Phi(t,r)\propto t^{-5/6}\,\sin(\mu t+\varphi).
\end{equation}
Here we see that the decay law at sufficiently late times $t \sim 1500$ for $\mu=10$, $M=1.66$, the decay law can be approximated as  
\begin{equation}\label{asymptotictail2}
\Phi(t,r)\propto t^{-7/8}\,\sin(\mu t+\varphi),
\end{equation}
as shown in fig. \ref{fig:TD2}. However, this behavior has been inferred only from a numerical fit and may not yet represent the true asymptotic regime.
The precise law for the asymptotic decay could be found either by fitting extensive numerical data for various values of $\gamma$ and $\ell$ or via analysis of Green functions.

Notice that the intermediate late-time tails, which appear earlier for larger values of the product $\mu M$, contaminate the late-time ringdown signal, making it difficult to achieve truly good agreement with the WKB results, as illustrated in Fig.~\ref{fig:TD3}. Although higher-order WKB methods have frequently been employed to compute quasinormal modes of massive fields \cite{Konoplya:2002wt,Konoplya:2018qov,Lutfuoglu:2026fpx,Lutfuoglu:2026gis,Lutfuoglu:2026xlo,Bolokhov:2026dfg,Bolokhov:2026uol,Skvortsova:2026jtx}, with the difference between nearby orders often used as an indicator of stability, the WKB series is only asymptotic and does not provide a mathematically rigorous error estimate. Therefore, more accurate approaches, such as the pseudospectral method \cite{Fortuna:2020obg,Jansen:2017oag,Konoplya:2022zav,Batic:2025bkk,Batic:2026jdt} or the Leaver method \cite{Leaver:1985ax} (see examples for various spacetimes in \cite{Batic:2025xge,Kanti:2006ua,Konoplya:2020hyk,Saka:2025xxl,Zinhailo:2024kbq}), are required in order to reliably approach the quasi-resonant regime.

Our confidence in the existence of quasi-resonances is based not solely on the apparent convergence of the WKB approximation, but primarily on the observed tendency of the frequencies to approach their Schwarzschild counterparts as the field mass increases. For the Schwarzschild black hole, the existence of quasi-resonances is well established through highly accurate methods, and the corresponding modes are known to asymptotically approach the quasi-resonant regime.

\begin{figure*}
\resizebox{\linewidth}{!}{
\includegraphics{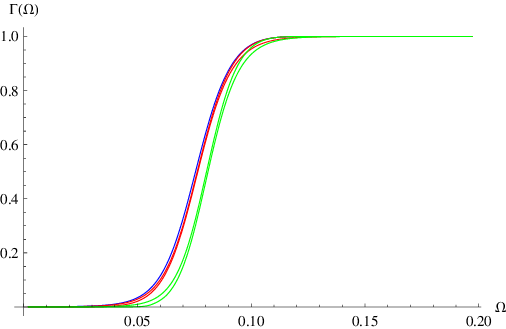}
\includegraphics{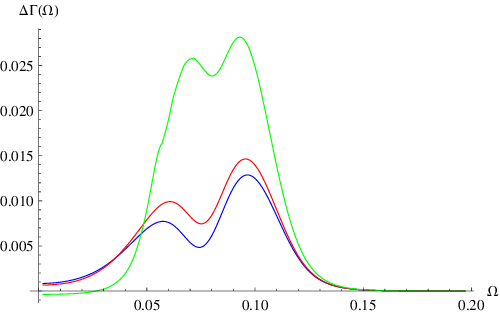}
}
\caption{Grey-body factors calculated by the 6th order WKB technique and by the correspondence with QNMs (left). The difference between the results obtained by the two methods (right). Here $M=4$, $\gamma=0.1$, $\ell=1$, $\mu=0$ (blue), $\mu=0.02$ (red), $\mu=0.05$ (green).}\label{fig:GBL1}
\end{figure*}

\begin{figure*}
\resizebox{\linewidth}{!}{
\includegraphics{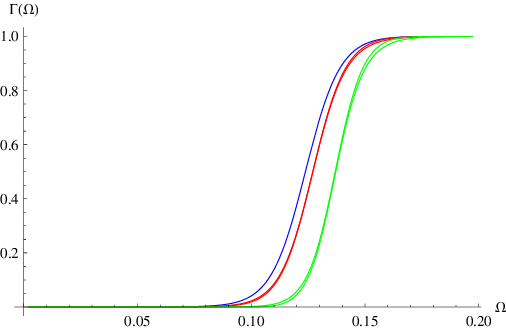}
\includegraphics{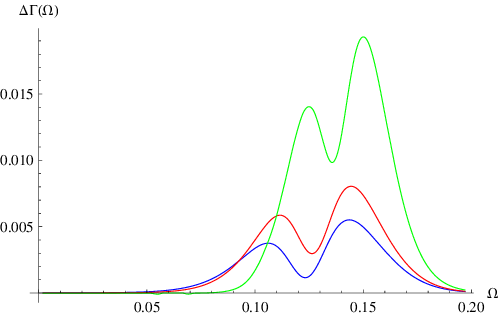}
}
\caption{Grey-body factors calculated by the 6th order WKB technique and by the correspondence with QNMs (left). The difference between the results obtained by the two methods (right). Here $M=4$, $\gamma=0.1$, $\ell=2$, $\mu=0$ (blue), $\mu=0.05$ (red), $\mu=0.1$ (green).}\label{fig:GBL2}
\end{figure*}

\begin{figure*}
\resizebox{\linewidth}{!}{
\includegraphics{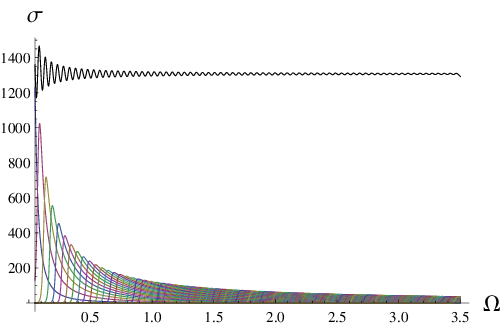}
\includegraphics{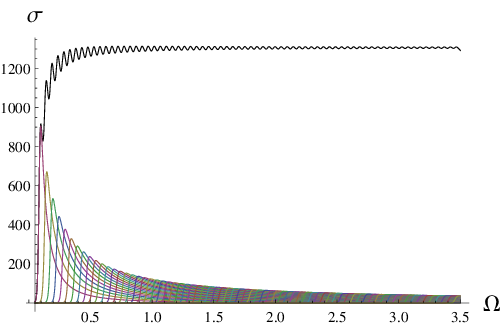}
}
\caption{Absorption cross-section for the first 70  multipole numbers together with the total cross-section for $M=4$, $\gamma=0.1$, $\mu=0$ (left) and $\mu=0.07$ (right).}\label{fig:SIGMA}
\end{figure*}

\section{Grey-body factors and absorption cross-sections}
\label{sec:GBF}

When a quantum field is emitted near a black hole horizon,
its propagation toward infinity is partially hindered by the curvature-induced
effective potential outside the horizon.  
As a result, only a fraction of the produced radiation escapes to infinity,
while the rest is reflected back and reabsorbed by the black hole.  
This filtering of the Hawking spectrum is quantified by
the grey-body factors, which encode the transmission probability
for each partial wave and thus determine the observed deviation from a purely
thermal spectrum.

\subsection{Definition and boundary conditions}\label{sec:gbf}

For each multipole number $\ell$,
the stationary wave equation~\eqref{eq:RadialEq} can be treated as a
one-dimensional scattering problem in the tortoise coordinate $r_*$:
\begin{equation}
\frac{d^{2}\Psi_\ell}{dr_*^{2}}+\bigl(\Omega^{2}-V_\ell(r)\bigr)\Psi_\ell=0.
\label{eq:waveGBF}
\end{equation}
The physically relevant boundary conditions correspond to
a unit-amplitude incoming wave from infinity,
a reflected component there, and a purely ingoing wave at the event horizon:
\begin{equation}
\Psi_\ell(r) =
\begin{cases}
T_\ell(\Omega)\,e^{-i\Omega r_*}, & r_* \to -\infty \quad (r\to r_h),\\[6pt]
e^{-i\Omega r_*} + R_\ell(\Omega)\,e^{+i\Omega r_*}, & r_* \to +\infty,
\end{cases}
\label{eq:BC_GBF}
\end{equation}
where $R_\ell(\Omega)$ and $T_\ell(\Omega)$ are the complex reflection and
transmission amplitudes, respectively.
The normalization condition ensuring energy conservation is
\begin{equation}
|R_\ell|^{2} + |T_\ell|^{2} = 1.
\end{equation}

The grey-body factor for a given mode is then defined as
\begin{equation}
\Gamma_\ell(\Omega) = |T_\ell(\Omega)|^2 = 1 - |R_\ell(\Omega)|^2,
\label{eq:GBF_def}
\end{equation}
representing the probability for an incoming wave to be transmitted through
the effective potential barrier.

The WKB method provides also an efficient
semi-analytic way to estimate $\Gamma_\ell(\Omega)$.
Following the formulation of Schutz and Will~\cite{Schutz:1985km}
and its higher-order extensions~\cite{Iyer:1986np,Konoplya:2003ii,Matyjasek:2017psv,Konoplya:2019hlu},
the transmission probability at the $N$-th WKB order reads
\begin{eqnarray}
&&\Gamma_\ell(\Omega) = 
\frac{1}{1 + \exp\!\bigl(2\pi\,K_\ell(\Omega)\bigr)},
\nonumber\\
&&K_\ell(\Omega) =
\frac{i\bigl(\Omega^{2}-V_0\bigr)}{\sqrt{-2\,V_0^{(2)}}}
+ \sum_{k=2}^{N}\Lambda_k\!\left(\{V_0^{(j)}\},\Omega\right),
\label{eq:GBF_WKB}
\end{eqnarray}
where $V_0$ and $V_0^{(2)}$ denote the value and second derivative of the effective potential at its maximum, while $\Lambda_k$ are known correction terms involving higher derivatives of $V(r)$ evaluated at the same point.
The WKB expression~\eqref{eq:GBF_WKB} smoothly interpolates between two limits:
$\Gamma_\ell \to 1$ for $\Omega^2 \gg V_0$ (wave passes over the barrier)
and $\Gamma_\ell \to 0$ for $\Omega^2 \ll V_0$ (wave is reflected),
thereby offering a transparent physical interpretation of transmission suppression.
For accurate calculations of grey-body factors at small values of the multipole moment, the numerical methods could be used \cite{Arbey:2025dnc, Page:1976df, Page:1976ki, Konoplya:2025ixm, Calza:2025whq}.

The grey-body factors for various values of the parameters calculated with the help of the 6th order WKB method as a function of the real frequency $\Omega$ are shown in figs. \ref{fig:GBL1} and \ref{fig:GBL2}. For larger values of $\ell$, the grey-body factors are strongly suppressed shifting the radiation to higher frequencies, which is a common and well-known behavior observed in numerous studies \cite{Lutfuoglu:2025mqa,Lutfuoglu:2025kqp,Lutfuoglu:2026gey,Konoplya:2010vz,Lutfuoglu:2025eik,Lutfuoglu:2026uzy,Dubinsky:2024vbn,Bolokhov:2024voa,Malik:2025qnr,Malik:2024wvs} and is associated with the increasing height of the effective potential barrier.  As illustrated in figs.~\ref{fig:GBL1} and \ref{fig:GBL2}, increasing the field mass $\mu$ leads to a pronounced suppression of the grey-body factors at low frequencies. Physically, this behavior can be traced to the structure of the effective potential shown in figs.~\ref{fig:Potential1} and \ref{fig:Potential2}: a heavier field generates a higher potential barrier, thereby reducing the transmission probability through it and diminishing the emitted flux of low-frequency quanta.

The total absorption cross-section is obtained by summing the contributions
from all partial waves:
\begin{equation}
\sigma_{\text{abs}}(\Omega)
= \frac{\pi}{\Omega^{2}}
\sum_{\ell=0}^{\infty} (2\ell+1)\,\Gamma_\ell(\Omega).
\label{eq:sigma_abs}
\end{equation}
Physically, $\sigma_{\text{abs}}$ represents the effective area of the black hole
that interacts with the incoming flux at frequency $\omega$.
At low frequencies, only the lowest multipoles contribute,
and the cross-section decreases rapidly with $\Omega$.
In the high-frequency limit, $\sigma_{\text{abs}}$ tends to the geometric-optics
capture cross-section $\sigma_{\text{geo}} = \pi b_c^2$,
where $b_c$ is the critical impact parameter for unstable photon orbits.
The grey-body factors thus bridge the microscopic scattering properties of the field
and the macroscopic observables, such as energy emission and absorption rates (see fig. \ref{fig:SIGMA}).

The spectrum of quasinormal modes obtained for the Bonanno–Reuter black hole can also be used to estimate the grey-body factors of massive fields through the approximate correspondence between transmission probabilities and quasinormal frequencies~\cite{Konoplya:2024lir}. In its simplest form, this relation can be expressed as
\[\begin{array}{r}
\Gamma_{\ell}(\Omega) =
\left[
1 + 
\exp\!\left(
\dfrac{2\pi\bigl[\Omega^{2} - \mathrm{Re}(\omega_{0})^{2}\bigr]}
     {4\,\mathrm{Re}(\omega_{0})\,\mathrm{Im}(\omega_{0})}
\right)
\right]^{-1}\\
+ \text{higher order terms},
\end{array}
\]
where $\omega_{0}$ denotes the fundamental quasinormal frequency, while higher-order corrections originate mainly from the first overtone.

This correspondence is known to be quantitatively accurate in the eikonal regime, where the WKB expansion performs well, but it becomes only approximate for the lowest multipoles, as shown in several recent analyses~\cite{Han:2025cal,Dubinsky:2025nxv,Malik:2024cgb,Lutfuoglu:2025blw,Lutfuoglu:2025ldc,Bolokhov:2024otn,Malik:2025erb}.  {However, the correspondence is known to be broken when the WKB approach on which it is based is invalid or incompletely describe the spectrum of quasinormal modes. For example, the correspondence breaks down in some theories with higher curvature corrections \cite{Konoplya:2017wot,Bolokhov:2023dxq} or when the effective potential has double-well \cite{Konoplya:2025hgp}. In case of asymptotically de Sitter black holes, the WKB method does not reproduce the de Sitter branch of modes \cite{Konoplya:2022gjp} and consequently the correspondence relate only the Schwarzschild branch of modes with the grey-body factors \cite{Malik:2024cgb}.  Here, following recent works \cite{Konoplya:2023moy,Matyjasek:2026yiu}, we employ the correspondence formula extended to second order beyond the eikonal limit.

For massive fields in asymptotically flat backgrounds, the validity of the relation is further constrained: it holds only when the dimensionless parameter $\mu M$ is small enough for the WKB method to remain reliable. Once the field mass grows, the effective potential deviates substantially from the single-barrier form assumed by the WKB approximation, causing the correspondence to deteriorate.

\section{Conclusions}
\label{sec:conclusions}

We have studied quasinormal modes, grey-body factors, and absorption cross-sections for a massive scalar field propagating in the background of the Bonanno--Reuter quantum-corrected black hole. This geometry, derived from the renormalization-group improvement of the Schwarzschild metric in asymptotically safe gravity, replaces the central singularity with a regular de~Sitter core and introduces $1/r^{3}$ quantum corrections at large distances.

Our results show that increasing the mass of the field $\mu$ leads to a strong suppression of the damping rate and a systematic decrease of the real oscillation frequency. Extrapolation to larger $\mu$ indicates the onset of quasi-resonances --- extremely long-lived oscillations analogous to those found for massive fields in the classical Schwarzschild background. The WKB–Padé method and time-domain integration yield consistent results within a relative deviation of less than one percent in their common range of applicability, confirming the robustness of our numerical analysis.

At late times, the time-domain profiles exhibit oscillatory tails with a power-law envelope $$\Phi \propto t^{-7/8}\sin(\mu t + \varphi),$$ distinct from the classical massless case. However, in order to be sure that this is a true asymptotic regime, analytical treatment is necessary to confirm this observation. The grey-body factors, computed via the WKB transmission formalism and verified through their correspondence with quasinormal data, demonstrate a pronounced dependence on the field’s mass: larger $\mu$ significantly suppresses low-frequency transmission and shifts the effective emission spectrum toward higher frequencies. This suppression directly reflects the increased height of the effective potential barrier caused by the mass term. The total absorption cross-section, obtained from the sum of partial multipole contributions, approaches the geometric-optics limit at high frequencies, while at low frequencies it is diminished by both quantum corrections and field mass.

In the present work, our attention has been restricted to the fundamental quasinormal mode and, where applicable, to the first overtone — those parts of the spectrum for which the WKB method provides dependable accuracy. Nevertheless, the higher-overtone sector remains of considerable theoretical interest. Since these modes are highly sensisite to the changes of the metric in the vicinity of the event horizon~\cite{Konoplya:2022pbc,Stuchlik:2025ezz}, they are expected to be especially sensitive to the details of quantum modifications of the geometry and may exhibit a complex interplay with the quasi-resonant regime characteristic of massive fields. Capturing their behavior requires techniques that go beyond the WKB expansion or time-domain integration, such as the convergent Frobenius (Leaver) approach~\cite{Leaver:1985ax} or pseudo-spectral methods. A thorough investigation of this regime  is deferred to future research.

\begin{acknowledgments}
The author would like to thank R. A. Konoplya for stimulating discussions and Marco Calza for useful correspondence. 
\end{acknowledgments}

\bibliography{bibliography}

@article{Malik:2025ava,
    author = "Malik, Zainab",
    title = {{Long-lived quasinormal modes of brane-localized Reissner{\textendash}Nordstr{\"o}m{\textendash}de Sitter black holes}},
    eprint = "2504.12570",
    archivePrefix = "arXiv",
    primaryClass = "gr-qc",
    doi = "10.1016/j.aop.2025.170238",
    journal = "Annals Phys.",
    volume = "482",
    pages = "170238",
    year = "2025"
}

@article{Han:2025cal,
    author = "Han, Hyewon and Gwak, Bogeun",
    title = "{Correspondence between quasinormal modes and greybody factors in five-dimensional black holes}",
    eprint = "2508.12989",
    archivePrefix = "arXiv",
    primaryClass = "gr-qc",
    doi = "10.1103/n2ns-drkp",
    journal = "Phys. Rev. D",
    volume = "113",
    number = "6",
    pages = "064058",
    year = "2026"
}

@article{Bronnikov:2021liv,
    author = "Bronnikov, Kirill A. and Konoplya, Roman A. and Pappas, Thomas D.",
    title = "{General parametrization of wormhole spacetimes and its application to shadows and quasinormal modes}",
    eprint = "2102.10679",
    archivePrefix = "arXiv",
    primaryClass = "gr-qc",
    doi = "10.1103/PhysRevD.103.124062",
    journal = "Phys. Rev. D",
    volume = "103",
    number = "12",
    pages = "124062",
    year = "2021"
}

@article{Dubinsky:2025nxv,
    author = "Dubinsky, Alexey",
    title = "{Gravitational perturbations of Dymnikova black holes: Grey-body factors and absorption cross-sections}",
    eprint = "2509.11017",
    archivePrefix = "arXiv",
    primaryClass = "gr-qc",
    doi = "10.1016/j.aop.2025.170299",
    journal = "Annals Phys.",
    volume = "485",
    pages = "170299",
    year = "2026"
}

@article{Konoplya:2020fwg,
    author = "Konoplya, R. A.",
    title = "{Conformal Weyl gravity via two stages of quasinormal ringing and late-time behavior}",
    eprint = "2012.13020",
    archivePrefix = "arXiv",
    primaryClass = "gr-qc",
    doi = "10.1103/PhysRevD.103.044033",
    journal = "Phys. Rev. D",
    volume = "103",
    number = "4",
    pages = "044033",
    year = "2021"
}

@article{Lutfuoglu:2025pzi,
    author = {L{\"u}tf{\"u}o{\u{g}}lu, Bekir Can and Saka, Erdin{\c{c}} Ula{\c{s}} and Shermatov, Abubakir and Rayimbaev, Javlon and Ibragimov, Inomjon and Muminov, Sokhibjan},
    title = "{Gravitational quasinormal modes of Dymnikova black holes}",
    eprint = "2509.24633",
    archivePrefix = "arXiv",
    primaryClass = "gr-qc",
    doi = "10.1016/j.aop.2026.170360",
    journal = "Annals Phys.",
    volume = "487",
    pages = "170360",
    year = "2026"
}

@article{Bolokhov:2025egl,
    author = "Bolokhov, S. V. and Skvortsova, Milena",
    title = "{Gravitational quasinormal modes of the Hayward spacetime}",
    eprint = "2508.19989",
    archivePrefix = "arXiv",
    primaryClass = "gr-qc",
    month = "8",
    year = "2025"
}

@article{LIGOScientific:2016aoc,
    author = "Abbott, B. P. and others",
    collaboration = "LIGO Scientific, Virgo",
    title = "{Observation of Gravitational Waves from a Binary Black Hole Merger}",
    eprint = "1602.03837",
    archivePrefix = "arXiv",
    primaryClass = "gr-qc",
    reportNumber = "LIGO-P150914",
    doi = "10.1103/PhysRevLett.116.061102",
    journal = "Phys. Rev. Lett.",
    volume = "116",
    number = "6",
    pages = "061102",
    year = "2016"
}

@article{LIGOScientific:2017vwq,
    author = "Abbott, B. P. and others",
    collaboration = "LIGO Scientific, Virgo",
    title = "{GW170817: Observation of Gravitational Waves from a Binary Neutron Star Inspiral}",
    eprint = "1710.05832",
    archivePrefix = "arXiv",
    primaryClass = "gr-qc",
    reportNumber = "LIGO-P170817",
    doi = "10.1103/PhysRevLett.119.161101",
    journal = "Phys. Rev. Lett.",
    volume = "119",
    number = "16",
    pages = "161101",
    year = "2017"
}

@article{LIGOScientific:2020zkf,
    author = "Abbott, R. and others",
    collaboration = "LIGO Scientific, Virgo",
    title = "{GW190814: Gravitational Waves from the Coalescence of a 23 Solar Mass Black Hole with a 2.6 Solar Mass Compact Object}",
    eprint = "2006.12611",
    archivePrefix = "arXiv",
    primaryClass = "astro-ph.HE",
    reportNumber = "LIGO-P190814",
    doi = "10.3847/2041-8213/ab960f",
    journal = "Astrophys. J. Lett.",
    volume = "896",
    number = "2",
    pages = "L44",
    year = "2020"
}

@article{Macedo:2024dqb,
    author = "Mac{\^e}do, M. H. and Furtado, J. and Alencar, G. and Landim, R. R.",
    title = "{Thermodynamics and quasinormal modes of the Dymnikova black hole in higher dimensions}",
    eprint = "2404.02818",
    archivePrefix = "arXiv",
    primaryClass = "gr-qc",
    doi = "10.1016/j.aop.2024.169833",
    journal = "Annals Phys.",
    volume = "471",
    pages = "169833",
    year = "2024"
}

@article{Leaver:1985ax,
    author = "Leaver, E. W.",
    title = "{An Analytic representation for the quasi normal modes of Kerr black holes}",
    doi = "10.1098/rspa.1985.0119",
    journal = "Proc. Roy. Soc. Lond. A",
    volume = "402",
    pages = "285--298",
    year = "1985"
}

@article{Malik:2025dxn,
    author = "Malik, Zainab",
    title = "{Gravitational Perturbations of the Hayward Spacetime and Testing the Correspondence between Quasinormal Modes and Grey-body Factors}",
    eprint = "2508.19178",
    archivePrefix = "arXiv",
    primaryClass = "gr-qc",
    doi = "10.1007/s10773-025-06198-w",
    journal = "Int. J. Theor. Phys.",
    volume = "64",
    number = "11",
    pages = "314",
    year = "2025"
}

@article{Bolokhov:2025lnt,
    author = "Bolokhov, S. V. and Skvortsova, Milena",
    title = "{Gravitational Quasinormal Modes and Grey-Body Factors of Bonanno-Reuter Regular Black Holes}",
    eprint = "2507.07196",
    archivePrefix = "arXiv",
    primaryClass = "gr-qc",
    journal = "Int. J. Grav. Theor. Phys.",
    volume = "1",
    pages = "3",
    year = "2025"
}

@article{KAGRA:2013rdx,
    author = "Abbott, B. P. and others",
    collaboration = "KAGRA, LIGO Scientific, Virgo",
    title = "{Prospects for observing and localizing gravitational-wave transients with Advanced LIGO, Advanced Virgo and KAGRA}",
    eprint = "1304.0670",
    archivePrefix = "arXiv",
    primaryClass = "gr-qc",
    reportNumber = "LIGO-P1200087, VIR-0288A-12, JGW-P1808427",
    doi = "10.1007/s41114-020-00026-9",
    journal = "Living Rev. Rel.",
    volume = "19",
    pages = "1",
    year = "2016"
}

@article{Bolokhov:2023dxq,
    author = "Bolokhov, S. V.",
    title = "{Black holes in Starobinsky-Bel-Robinson Gravity and the breakdown of quasinormal modes/null geodesics correspondence}",
    eprint = "2310.12326",
    archivePrefix = "arXiv",
    primaryClass = "gr-qc",
    doi = "10.1016/j.physletb.2024.138879",
    journal = "Phys. Lett. B",
    volume = "856",
    pages = "138879",
    year = "2024"
}

@article{Konoplya:2022pbc,
    author = "Konoplya, R. A. and Zhidenko, A.",
    title = "{First few overtones probe the event horizon geometry}",
    eprint = "2209.00679",
    archivePrefix = "arXiv",
    primaryClass = "gr-qc",
    doi = "10.1016/j.jheap.2024.10.015",
    journal = "JHEAp",
    volume = "44",
    pages = "419--426",
    year = "2024"
}

@article{Konoplya:2020jgt,
    author = "Konoplya, R. A. and Zinhailo, A. F. and Stuchlik, Z.",
    title = "{Quasinormal modes and Hawking radiation of black holes in cubic gravity}",
    eprint = "2006.10462",
    archivePrefix = "arXiv",
    primaryClass = "gr-qc",
    doi = "10.1103/PhysRevD.102.044023",
    journal = "Phys. Rev. D",
    volume = "102",
    number = "4",
    pages = "044023",
    year = "2020"
}

@article{Dubinsky:2024nzo,
    author = "Dubinsky, Alexey and Zinhailo, Antonina F.",
    title = "{Analytic expressions for grey-body factors of the general parametrized spherically symmetric black holes}",
    eprint = "2410.15232",
    archivePrefix = "arXiv",
    primaryClass = "gr-qc",
    doi = "10.1209/0295-5075/adbc17",
    journal = "EPL",
    volume = "149",
    number = "6",
    pages = "69004",
    year = "2025"
}

@article{Malik:2025qnr,
    author = "Malik, Zainab",
    title = "{Bonanno-Reuter regular black hole: quasi-resonances, grey-body factors and absorption cross-sections of a massive scalar field}",
    eprint = "2510.06689",
    archivePrefix = "arXiv",
    primaryClass = "gr-qc",
    month = "10",
    year = "2025"
}

@article{Matyjasek:2026yiu,
    author = "Matyjasek, Jerzy and Konoplya, Roman A. and Zhidenko, Alexander",
    title = "{An Efficient Higher-Order WKB Code for Quasinormal Modes and Greybody Factors}",
    eprint = "2603.12466",
    archivePrefix = "arXiv",
    primaryClass = "gr-qc",
    doi = "10.53941/ijgtp.2026.100005",
    journal = "Int. J. Grav. Theor. Phys.",
    volume = "2",
    number = "1",
    pages = "5",
    year = "2026"
}

@article{Lutfuoglu:2025blw,
    author = {L{\"u}tf{\"u}o{\u{g}}lu, Bekir Can and Saka, Erdin{\c{c}} Ula{\c{s}} and Shermatov, Abubakir and Rayimbaev, Javlon and Ibragimov, Inomjon and Muminov, Sokhibjan},
    title = "{Proper-time approach in asymptotic safety via black hole quasinormal modes and grey-body factors}",
    eprint = "2509.15923",
    archivePrefix = "arXiv",
    primaryClass = "gr-qc",
    doi = "10.1140/epjc/s10052-025-14950-z",
    journal = "Eur. Phys. J. C",
    volume = "85",
    number = "10",
    pages = "1190",
    year = "2025"
}

@article{Dubinsky:2024vbn,
    author = "Dubinsky, Alexey",
    title = "{Gray-body factors for gravitational and electromagnetic perturbations around Gibbons{\textendash}Maeda{\textendash}Garfinkle{\textendash}Horowitz{\textendash}Strominger black holes}",
    eprint = "2412.00625",
    archivePrefix = "arXiv",
    primaryClass = "gr-qc",
    doi = "10.1142/S0217732325501111",
    journal = "Mod. Phys. Lett. A",
    volume = "40",
    number = "28",
    pages = "2550111",
    year = "2025"
}

@article{Lutfuoglu:2026uzy,
    author = {L{\"u}tf{\"u}o{\u{g}}lu, Bekir Can},
    title = "{Quasinormal Modes of a Massive Scalar Field in 4D Einstein--Gauss--Bonnet Black Hole Spacetimes}",
    eprint = "2603.24424",
    archivePrefix = "arXiv",
    primaryClass = "gr-qc",
    month = "3",
    year = "2026"
}

@article{Konoplya:2017wot,
    author = "Konoplya, R. A. and Stuchl{\'\i}k, Z.",
    title = "{Are eikonal quasinormal modes linked to the unstable circular null geodesics?}",
    eprint = "1705.05928",
    archivePrefix = "arXiv",
    primaryClass = "gr-qc",
    doi = "10.1016/j.physletb.2017.06.015",
    journal = "Phys. Lett. B",
    volume = "771",
    pages = "597--602",
    year = "2017"
}

@article{Konoplya:2022gjp,
    author = "Konoplya, R. A.",
    title = "{Further clarification on quasinormal modes/circular null geodesics correspondence}",
    eprint = "2210.08373",
    archivePrefix = "arXiv",
    primaryClass = "gr-qc",
    doi = "10.1016/j.physletb.2023.137674",
    journal = "Phys. Lett. B",
    volume = "838",
    pages = "137674",
    year = "2023"
}

@article{Konoplya:2025hgp,
    author = "Konoplya, R. A. and Stashko, O. S.",
    title = "{Transition from regular black holes to wormholes in covariant effective quantum gravity: Scattering, quasinormal modes, and Hawking radiation}",
    eprint = "2502.05689",
    archivePrefix = "arXiv",
    primaryClass = "gr-qc",
    doi = "10.1103/PhysRevD.111.084031",
    journal = "Phys. Rev. D",
    volume = "111",
    number = "8",
    pages = "084031",
    year = "2025"
}

@article{Bolokhov:2025uxz,
    author = "Bolokhov, S. V. and Skvortsova, Milena",
    title = "{Review of analytic results on quasinormal modes of black holes}",
    doi = "10.1134/S0202289325700306",
    journal = "Grav. Cosmol.",
    volume = "31",
    number = "4",
    pages = "423--446",
    eprint = "2504.05014",
    archivePrefix = "arXiv",
    primaryClass = "gr-qc",
    year = "2025"
}

@article{Jansen:2017oag,
    author = "Jansen, Aron",
    title = "{Overdamped modes in Schwarzschild-de Sitter and a Mathematica package for the numerical computation of quasinormal modes}",
    eprint = "1709.09178",
    archivePrefix = "arXiv",
    primaryClass = "gr-qc",
    doi = "10.1140/epjp/i2017-11825-9",
    journal = "Eur. Phys. J. Plus",
    volume = "132",
    number = "12",
    pages = "546",
    year = "2017"
}

@article{Fortuna:2020obg,
    author = "Fortuna, Sean and Vega, Ian",
    title = "{Bernstein spectral method for quasinormal modes and other eigenvalue problems}",
    eprint = "2003.06232",
    archivePrefix = "arXiv",
    primaryClass = "gr-qc",
    doi = "10.1140/epjc/s10052-023-12350-9",
    journal = "Eur. Phys. J. C",
    volume = "83",
    number = "12",
    pages = "1170",
    year = "2023"
}

@article{Konoplya:2002wt,
    author = "Konoplya, R. A.",
    title = "{Massive charged scalar field in a Reissner-Nordstrom black hole background: Quasinormal ringing}",
    eprint = "gr-qc/0210105",
    archivePrefix = "arXiv",
    doi = "10.1016/S0370-2693(02)02974-X",
    journal = "Phys. Lett. B",
    volume = "550",
    pages = "117--120",
    year = "2002"
}

@article{Bolokhov:2024bke,
    author = "Bolokhov, Sergey",
    title = "{Long lived quasinormal modes in the effective quantum gravity}",
    doi = "10.1140/epjc/s10052-025-14883-7",
    journal = "Eur. Phys. J. C",
    volume = "85",
    number = "10",
    pages = "1166",
    year = "2025"
}

@article{Konoplya:2011qq,
    author = "Konoplya, R. A. and Zhidenko, A.",
    title = "{Quasinormal modes of black holes: From astrophysics to string theory}",
    eprint = "1102.4014",
    archivePrefix = "arXiv",
    primaryClass = "gr-qc",
    doi = "10.1103/RevModPhys.83.793",
    journal = "Rev. Mod. Phys.",
    volume = "83",
    pages = "793--836",
    year = "2011"
}

@article{Berti:2009kk,
    author = "Berti, Emanuele and Cardoso, Vitor and Starinets, Andrei O.",
    title = "{Quasinormal modes of black holes and black branes}",
    eprint = "0905.2975",
    archivePrefix = "arXiv",
    primaryClass = "gr-qc",
    doi = "10.1088/0264-9381/26/16/163001",
    journal = "Class. Quant. Grav.",
    volume = "26",
    pages = "163001",
    year = "2009"
}

@article{Skvortsova:2024msa,
    author = "Skvortsova, Milena",
    title = "{Quantum corrected black holes: testing the correspondence between grey-body factors and quasinormal modes}",
    eprint = "2411.06007",
    archivePrefix = "arXiv",
    primaryClass = "gr-qc",
    doi = "10.1140/epjc/s10052-025-14589-w",
    journal = "Eur. Phys. J. C",
    volume = "85",
    number = "8",
    pages = "854",
    year = "2025"
}

@article{Churilova:2019qph,
    author = "Churilova, M. S. and Konoplya, R. A. and Zhidenko, A.",
    title = "{Arbitrarily long-lived quasinormal modes in a wormhole background}",
    eprint = "1911.05246",
    archivePrefix = "arXiv",
    primaryClass = "gr-qc",
    doi = "10.1016/j.physletb.2020.135207",
    journal = "Phys. Lett. B",
    volume = "802",
    pages = "135207",
    year = "2020"
}

@article{Konoplya:2013sba,
    author = "Konoplya, R. A. and Zhidenko, A.",
    title = "{Instability of D-dimensional extremally charged Reissner-Nordstrom(-de Sitter) black holes: Extrapolation to arbitrary D}",
    eprint = "1309.7667",
    archivePrefix = "arXiv",
    primaryClass = "hep-th",
    doi = "10.1103/PhysRevD.89.024011",
    journal = "Phys. Rev. D",
    volume = "89",
    number = "2",
    pages = "024011",
    year = "2014"
}

@article{Qian:2022kaq,
    author = "Qian, Wei-Liang and Lin, Kai and Shao, Cai-Ying and Wang, Bin and Yue, Rui-Hong",
    title = "{On the late-time tails of massive perturbations in spherically symmetric black holes}",
    eprint = "2203.04477",
    archivePrefix = "arXiv",
    primaryClass = "gr-qc",
    doi = "10.1140/epjc/s10052-022-10910-z",
    journal = "Eur. Phys. J. C",
    volume = "82",
    number = "10",
    pages = "931",
    year = "2022"
}

@article{Varghese:2011ku,
    author = "Varghese, Nijo and Kuriakose, V. C.",
    title = "{Evolution of massive fields around a black hole in Horava gravity}",
    eprint = "1011.6608",
    archivePrefix = "arXiv",
    primaryClass = "gr-qc",
    doi = "10.1007/s10714-011-1201-y",
    journal = "Gen. Rel. Grav.",
    volume = "43",
    pages = "2757--2767",
    year = "2011"
}

@article{Malik:2024bmp,
    author = "Malik, Zainab",
    title = "{Quasinormal modes of the Mannheim{\textendash}Kazanas black holes}",
    doi = "10.1515/zna-2024-0153",
    journal = "Z. Naturforsch. A",
    volume = "79",
    number = "11",
    pages = "1063--1073",
    year = "2024"
}

@article{Dubinsky:2024jqi,
    author = "Dubinsky, Alexey",
    title = "{Telling late-time tails for a massive scalar field in the background of brane-localized black holes}",
    eprint = "2403.01883",
    archivePrefix = "arXiv",
    primaryClass = "gr-qc",
    doi = "10.1209/0295-5075/ad51a3",
    journal = "EPL",
    volume = "147",
    number = "1",
    pages = "19003",
    year = "2024"
}

@article{Cuyubamba:2016cug,
    author = "Cuyubamba, M. A. and Konoplya, R. A. and Zhidenko, A.",
    title = "{Quasinormal modes and a new instability of Einstein-Gauss-Bonnet black holes in the de Sitter world}",
    eprint = "1604.03604",
    archivePrefix = "arXiv",
    primaryClass = "gr-qc",
    doi = "10.1103/PhysRevD.93.104053",
    journal = "Phys. Rev. D",
    volume = "93",
    number = "10",
    pages = "104053",
    year = "2016"
}

@article{Bolokhov:2023ruj,
    author = "Bolokhov, S. V.",
    title = "{Long-lived quasinormal modes and oscillatory tails of the Bardeen spacetime}",
    doi = "10.1103/PhysRevD.109.064017",
    journal = "Phys. Rev. D",
    volume = "109",
    number = "6",
    pages = "064017",
    year = "2024"
}

@article{Chen:2023wkq,
    author = "Chen, H. and Sathiyaraj, T. and Hassanabadi, H. and Yang, Y. and Long, Z. W. and Tu, F. Q.",
    title = "{Quasinormal modes of the EGUP-corrected Schwarzschild black hole}",
    doi = "10.1007/s12648-023-02734-8",
    journal = "Indian J. Phys.",
    volume = "97",
    number = "14",
    pages = "4481--4489",
    year = "2023"
}

@article{Baruah:2023rhd,
    author = {Baruah, Anshuman and {\"O}vg{\"u}n, Ali and Deshamukhya, Atri},
    title = "{Quasinormal modes and bounding greybody factors of GUP-corrected black holes in Kalb{\textendash}Ramond gravity}",
    eprint = "2304.07761",
    archivePrefix = "arXiv",
    primaryClass = "gr-qc",
    doi = "10.1016/j.aop.2023.169393",
    journal = "Annals Phys.",
    volume = "455",
    pages = "169393",
    year = "2023"
}

@article{Moreira:2023cxy,
    author = "Moreira, Zeus S. and Lima Junior, Haroldo C. D. and Crispino, Lu{\'\i}s C. B. and Herdeiro, Carlos A. R.",
    title = "{Quasinormal modes of a holonomy corrected Schwarzschild black hole}",
    eprint = "2302.14722",
    archivePrefix = "arXiv",
    primaryClass = "gr-qc",
    doi = "10.1103/PhysRevD.107.104016",
    journal = "Phys. Rev. D",
    volume = "107",
    number = "10",
    pages = "104016",
    year = "2023"
}

@article{Fu:2023drp,
    author = "Fu, Guoyang and Zhang, Dan and Liu, Peng and Kuang, Xiao-Mei and Wu, Jian-Pin",
    title = "{Peculiar properties in quasinormal spectra from loop quantum gravity effect}",
    eprint = "2301.08421",
    archivePrefix = "arXiv",
    primaryClass = "gr-qc",
    doi = "10.1103/PhysRevD.109.026010",
    journal = "Phys. Rev. D",
    volume = "109",
    number = "2",
    pages = "026010",
    year = "2024"
}

@article{Heidari:2023ssx,
    author = "Heidari, N. and Hassanabadi, H. and Chen, H.",
    title = "{Quantum-corrected scattering of a Schwarzschild black hole with GUP effect}",
    doi = "10.1016/j.physletb.2023.137707",
    journal = "Phys. Lett. B",
    volume = "838",
    pages = "137707",
    year = "2023"
}

@article{Gundlach:1993tp,
    author = "Gundlach, Carsten and Price, Richard H. and Pullin, Jorge",
    title = "{Late time behavior of stellar collapse and explosions: 1. Linearized perturbations}",
    eprint = "gr-qc/9307009",
    archivePrefix = "arXiv",
    reportNumber = "NSF-ITP-93-84",
    doi = "10.1103/PhysRevD.49.883",
    journal = "Phys. Rev. D",
    volume = "49",
    pages = "883--889",
    year = "1994"
}

@article{Lutfuoglu:2025hwh,
    author = {L{\"u}tf{\"u}o{\u{g}}lu, B. C.},
    title = "{Long-lived quasinormal modes around regular black holes and wormholes in Covariant Effective Quantum Gravity}",
    eprint = "2504.09323",
    archivePrefix = "arXiv",
    primaryClass = "gr-qc",
    doi = "10.1088/1475-7516/2025/06/057",
    journal = "JCAP",
    volume = "06",
    pages = "057",
    year = "2025"
}

@article{Konoplya:2020der,
    author = "Konoplya, R. A. and Zhidenko, A.",
    title = "{4D Einstein-Lovelock black holes: Hierarchy of orders in curvature}",
    eprint = "2005.02225",
    archivePrefix = "arXiv",
    primaryClass = "gr-qc",
    doi = "10.1016/j.physletb.2020.135607",
    journal = "Phys. Lett. B",
    volume = "807",
    pages = "135607",
    year = "2020"
}

@article{Abdalla:2005hu,
    author = "Abdalla, E. and Konoplya, R. A. and Molina, C.",
    title = "{Scalar field evolution in Gauss-Bonnet black holes}",
    eprint = "hep-th/0507100",
    archivePrefix = "arXiv",
    doi = "10.1103/PhysRevD.72.084006",
    journal = "Phys. Rev. D",
    volume = "72",
    pages = "084006",
    year = "2005"
}

@article{Ishihara:2008re,
    author = "Ishihara, Hideki and Kimura, Masashi and Konoplya, Roman A. and Murata, Keiju and Soda, Jiro and Zhidenko, Alexander",
    title = "{Evolution of perturbations of squashed Kaluza-Klein black holes: escape from instability}",
    eprint = "0802.0655",
    archivePrefix = "arXiv",
    primaryClass = "hep-th",
    doi = "10.1103/PhysRevD.77.084019",
    journal = "Phys. Rev. D",
    volume = "77",
    pages = "084019",
    year = "2008"
}

@article{Konoplya:2024lch,
    author = "Konoplya, R. A. and Stashko, O. S.",
    title = "{Probing the effective quantum gravity via quasinormal modes and shadows of black holes}",
    eprint = "2408.02578",
    archivePrefix = "arXiv",
    primaryClass = "gr-qc",
    doi = "10.1103/PhysRevD.111.104055",
    journal = "Phys. Rev. D",
    volume = "111",
    number = "10",
    pages = "104055",
    year = "2025"
}

@article{Horowitz:2008bn,
    author = "Horowitz, Gary T. and Roberts, Matthew M.",
    title = "{Holographic Superconductors with Various Condensates}",
    eprint = "0810.1077",
    archivePrefix = "arXiv",
    primaryClass = "hep-th",
    doi = "10.1103/PhysRevD.78.126008",
    journal = "Phys. Rev. D",
    volume = "78",
    pages = "126008",
    year = "2008"
}

@article{Anacleto:2021qoe,
    author = "Anacleto, M. A. and Campos, J. A. V. and Brito, F. A. and Passos, E.",
    title = "{Quasinormal modes and shadow of a Schwarzschild black hole with GUP}",
    eprint = "2108.04998",
    archivePrefix = "arXiv",
    primaryClass = "gr-qc",
    doi = "10.1016/j.aop.2021.168662",
    journal = "Annals Phys.",
    volume = "434",
    pages = "168662",
    year = "2021"
}

@article{Anacleto:2019tdj,
    author = "Anacleto, M. A. and Brito, F. A. and Campos, J. A. V. and Passos, E.",
    title = "{Absorption and scattering of a noncommutative black hole}",
    eprint = "1907.13107",
    archivePrefix = "arXiv",
    primaryClass = "hep-th",
    doi = "10.1016/j.physletb.2020.135334",
    journal = "Phys. Lett. B",
    volume = "803",
    pages = "135334",
    year = "2020"
}

@article{Calza:2025whq,
    author = "Calz{\'a}, Marco",
    title = "{GrayHawk: A public code for calculating the Gray Body Factors of massless fields around spherically symmetric Black Holes}",
    eprint = "2502.04041",
    archivePrefix = "arXiv",
    primaryClass = "gr-qc",
    doi = "10.1016/j.dark.2025.101900",
    journal = "Phys. Dark Univ.",
    volume = "48",
    pages = "101900",
    year = "2025"
}

@article{Arbey:2025dnc,
    author = "Arbey, Alexandre and Calz{\`a}, Marco and Perez-Gonzalez, Yuber F.",
    title = "{Gray-body factors: Method matters}",
    eprint = "2502.17240",
    archivePrefix = "arXiv",
    primaryClass = "gr-qc",
    doi = "10.1016/j.dark.2025.101903",
    journal = "Phys. Dark Univ.",
    volume = "48",
    pages = "101903",
    year = "2025"
}

@article{Konoplya:2009hv,
    author = "Konoplya, R. A. and Zhidenko, A.",
    title = "{Holographic conductivity of zero temperature superconductors}",
    eprint = "0909.2138",
    archivePrefix = "arXiv",
    primaryClass = "hep-th",
    doi = "10.1016/j.physletb.2010.02.048",
    journal = "Phys. Lett. B",
    volume = "686",
    pages = "199--206",
    year = "2010"
}

@article{Herzog:2009xv,
    author = "Herzog, Christopher P.",
    title = "{Lectures on Holographic Superfluidity and Superconductivity}",
    eprint = "0904.1975",
    archivePrefix = "arXiv",
    primaryClass = "hep-th",
    reportNumber = "PUPT-2297",
    doi = "10.1088/1751-8113/42/34/343001",
    journal = "J. Phys. A",
    volume = "42",
    pages = "343001",
    year = "2009"
}

@article{Dubinsky:2024hmn,
    author = "Dubinsky, Alexey and Zinhailo, Antonina",
    title = "{Asymptotic decay and quasinormal frequencies of scalar and Dirac fields around dilaton-de Sitter black holes}",
    eprint = "2404.01834",
    archivePrefix = "arXiv",
    primaryClass = "gr-qc",
    doi = "10.1140/epjc/s10052-024-13206-6",
    journal = "Eur. Phys. J. C",
    volume = "84",
    number = "8",
    pages = "847",
    year = "2024"
}

@article{Dubinsky:2025bvf,
    author = "Dubinsky, Alexey",
    title = "{Long-Lived Quasinormal Modes and Quasi-Resonances around Non-Minimal Einstein-Yang-Mills Black Holes}",
	doi = {10.1140/epjc/s10052-025-14671-3},
    volume = "85",	
    number = {8},
	journal = {The European Physical Journal C},
	pages = {924},
    eprint = "2505.08545",
    archivePrefix = "arXiv",
    primaryClass = "gr-qc",
    month = "5",
    year = "2025"
}

@article{Dubinsky:2024rvf,
    author = "Dubinsky, Alexey",
    title = "{Analytic expressions for quasinormal modes of the general parametrized spherically symmetric black holes and the Hod's proposal}",
    eprint = "2409.16569",
    archivePrefix = "arXiv",
    primaryClass = "gr-qc",
    doi = "10.1016/j.physletb.2025.139251",
    journal = "Phys. Lett. B",
    volume = "861",
    pages = "139251",
    year = "2025"
}

@article{Lutfuoglu:2025mqa,
    author = {L{\"u}tf{\"u}o{\u{g}}lu, Bekir Can and Shermatov, Abubakir and Rayimbaev, Javlon and Matyoqubov, Muhammad and Sirajiddin, Otaboyev},
    title = "{Gravitational spectra and wave propagation in regular black holes supported by a Dehnen Halo}",
    eprint = "2511.22366",
    archivePrefix = "arXiv",
    primaryClass = "gr-qc",
    doi = "10.1140/epjc/s10052-025-15234-2",
    journal = "Eur. Phys. J. C",
    volume = "85",
    number = "12",
    pages = "1484",
    year = "2025"
}

@article{Lutfuoglu:2025eik,
    author = {L{\"u}tf{\"u}o{\u{g}}lu, Bekir Can},
    title = "{Grey-Body Factors and Absorption Cross-Sections~of Scalar and Dirac Fields in the Vicinity of Dilaton-De Sitter Black Hole}",
    eprint = "2510.10579",
    archivePrefix = "arXiv",
    primaryClass = "gr-qc",
    doi = "10.1002/prop.70074",
    journal = "Fortsch. Phys.",
    volume = "74",
    number = "1",
    pages = "e70074",
    year = "2026"
}

@article{Lutfuoglu:2025kqp,
    author = {L{\"u}tf{\"u}o{\u{g}}lu, B. C.},
    title = "{Long-lived quasinormal modes, grey-body factors and absorption cross-section of the black hole immersed in the Hernquist galactic halo}",
    eprint = "2510.25969",
    archivePrefix = "arXiv",
    primaryClass = "gr-qc",
    doi = "10.1016/j.physletb.2025.140082",
    journal = "Phys. Lett. B",
    volume = "872",
    pages = "140082",
    year = "2026"
}

@article{Lutfuoglu:2026gey,
    author = {L{\"u}tf{\"u}o{\u{g}}lu, Bekir Can and Rayimbaev, Javlon and Murodov, Sardor and Kurbanov, Jakhongir and Matyoqubov, Muhammad},
    title = "{A First-Order Eikonal Framework for Quasinormal Modes, Shadows, Strong Lensing, and Grey-Body Factors in a Scalarized Black-Hole Metric}",
    eprint = "2604.14999",
    archivePrefix = "arXiv",
    primaryClass = "gr-qc",
    month = "4",
    year = "2026"
}

@article{Konoplya:2010vz,
    author = "Konoplya, R. A. and Zhidenko, A.",
    title = "{Long life of Gauss-Bonnet corrected black holes}",
    eprint = "1004.3772",
    archivePrefix = "arXiv",
    primaryClass = "hep-th",
    doi = "10.1103/PhysRevD.82.084003",
    journal = "Phys. Rev. D",
    volume = "82",
    pages = "084003",
    year = "2010"
}

@article{Bonanno:2025dry,
    author = "Bonanno, Alfio M. and Konoplya, Roman A. and Oglialoro, Giovanni and Spina, Andrea",
    title = "{Regular black holes from proper-time flow in quantum gravity and their quasinormal modes, shadow and Hawking radiation}",
    eprint = "2509.12469",
    archivePrefix = "arXiv",
    primaryClass = "gr-qc",
    doi = "10.1088/1475-7516/2025/12/042",
    journal = "JCAP",
    volume = "12",
    pages = "042",
    year = "2025"
}

@article{Bonanno:2000ep,
    author = "Bonanno, Alfio and Reuter, Martin",
    title = "{Renormalization group improved black hole space-times}",
    eprint = "hep-th/0002196",
    archivePrefix = "arXiv",
    reportNumber = "INFN-CT-03-00, MZ-TH-00-04",
    doi = "10.1103/PhysRevD.62.043008",
    journal = "Phys. Rev. D",
    volume = "62",
    pages = "043008",
    year = "2000"
}

@article{Bolokhov:2024ixe,
    author = "Bolokhov, S. V.",
    title = "{Late time decay of scalar and Dirac fields around an asymptotically de Sitter black hole in the Euler{\textendash}Heisenberg electrodynamics}",
    eprint = "2404.09364",
    archivePrefix = "arXiv",
    primaryClass = "gr-qc",
    doi = "10.1140/epjc/s10052-024-12990-5",
    journal = "Eur. Phys. J. C",
    volume = "84",
    number = "6",
    pages = "634",
    year = "2024"
}

@article{Bolokhov:2024voa,
    author = "Bolokhov, S. V. and Konoplya, R. A.",
    title = "{Circumventing quantum gravity: Black holes evaporating into macroscopic wormholes}",
    eprint = "2410.10419",
    archivePrefix = "arXiv",
    primaryClass = "gr-qc",
    doi = "10.1103/PhysRevD.111.064007",
    journal = "Phys. Rev. D",
    volume = "111",
    number = "6",
    pages = "064007",
    year = "2025"
}

@article{Iyer:1986np,
    author = "Iyer, Sai and Will, Clifford M.",
    title = "{Black Hole Normal Modes: A {WKB} Approach. 1. Foundations and Application of a Higher Order {WKB} Analysis of Potential Barrier Scattering}",
    reportNumber = "Print-86-1482 (WASH. U., ST. LOUIS)",
    doi = "10.1103/PhysRevD.35.3621",
    journal = "Phys. Rev. D",
    volume = "35",
    pages = "3621",
    year = "1987"
}

@article{Konoplya:2003ii,
    author = "Konoplya, R. A.",
    title = "{Quasinormal behavior of the d-dimensional Schwarzschild black hole and higher order WKB approach}",
    eprint = "gr-qc/0303052",
    archivePrefix = "arXiv",
    doi = "10.1103/PhysRevD.68.024018",
    journal = "Phys. Rev. D",
    volume = "68",
    pages = "024018",
    year = "2003"
}

@article{Konoplya:2019hlu,
    author = "Konoplya, R. A. and Zhidenko, A. and Zinhailo, A. F.",
    title = "{Higher order WKB formula for quasinormal modes and grey-body factors: recipes for quick and accurate calculations}",
    eprint = "1904.10333",
    archivePrefix = "arXiv",
    primaryClass = "gr-qc",
    doi = "10.1088/1361-6382/ab2e25",
    journal = "Class. Quant. Grav.",
    volume = "36",
    pages = "155002",
    year = "2019"
}

@article{Skvortsova:2024wly,
    author = "Skvortsova, Milena",
    title = "{Ringing of Extreme Regular Black Holes}",
    eprint = "2405.15807",
    archivePrefix = "arXiv",
    primaryClass = "gr-qc",
    doi = "10.1134/S020228932470018X",
    journal = "Grav. Cosmol.",
    volume = "30",
    number = "3",
    pages = "279--288",
    year = "2024"
}

@article{Skvortsova:2024atk,
    author = "Skvortsova, Milena",
    title = "{Quasinormal Frequencies of Fields with Various Spin in the Quantum Oppenheimer{\textendash}Snyder Model of Black Holes}",
    eprint = "2405.06390",
    archivePrefix = "arXiv",
    primaryClass = "gr-qc",
    doi = "10.1002/prop.202400132",
    journal = "Fortsch. Phys.",
    volume = "72",
    number = "9-10",
    pages = "2400132",
    year = "2024"
}

@article{Bolokhov:2023bwm,
    author = "Bolokhov, S. V.",
    title = "{Long-lived quasinormal modes and overtones{\textquoteright} behavior of holonomy-corrected black holes}",
    eprint = "2311.05503",
    archivePrefix = "arXiv",
    primaryClass = "gr-qc",
    doi = "10.1103/PhysRevD.110.024010",
    journal = "Phys. Rev. D",
    volume = "110",
    number = "2",
    pages = "024010",
    year = "2024"
}

@article{Dubinsky:2025fwv,
    author = "Dubinsky, Alexey",
    title = "{Black Holes Immersed in Galactic Dark Matter Halo}",
    eprint = "2507.00256",
    archivePrefix = "arXiv",
    primaryClass = "gr-qc",
    journal = "International Journal of Gravitation and Theoretical Physics",
    volume = "1",
    pages = "2",
    year = "2025"
}

@article{Kokkotas:1999bd,
    author = "Kokkotas, Kostas D. and Schmidt, Bernd G.",
    title = "{Quasinormal modes of stars and black holes}",
    eprint = "gr-qc/9909058",
    archivePrefix = "arXiv",
    doi = "10.12942/lrr-1999-2",
    journal = "Living Rev. Rel.",
    volume = "2",
    pages = "2",
    year = "1999"
}

@article{Lutfuoglu:2025ldc,
    author = {L{\"u}tf{\"u}o{\u{g}}lu, Bekir Can},
    title = "{Black Holes in Proca-Gauss-Bonnet Gravity with Primary Hair: Particle Motion, Shadows, and Grey-Body Factors}",
    eprint = "2507.09246",
    archivePrefix = "arXiv",
    primaryClass = "gr-qc",
    journal = "Int.  J.  Grav.  Theor.  Phys",
    volume = "1",
    pages = "4",
    year = "2025"
}

@article{Malik:2024cgb,
    author = "Malik, Zainab",
    title = "{Correspondence between quasinormal modes and grey-body factors for massive fields in Schwarzschild-de~Sitter spacetime}",
    eprint = "2412.19443",
    archivePrefix = "arXiv",
    primaryClass = "gr-qc",
    doi = "10.1088/1475-7516/2025/04/042",
    journal = "JCAP",
    volume = "04",
    pages = "042",
    year = "2025"
}

@article{Konoplya:2025ixm,
    author = "Konoplya, Roman A. and Pappas, Thomas D.",
    title = "{Dirty black holes, clean signals: near-horizon vs.~environmental effects on grey-body factors and Hawking radiation}",
    eprint = "2507.01954",
    archivePrefix = "arXiv",
    primaryClass = "gr-qc",
    doi = "10.1088/1475-7516/2026/02/038",
    journal = "JCAP",
    volume = "02",
    pages = "038",
    year = "2026"
}

@article{Lutfuoglu:2025bsf,
    author = {L{\"u}tf{\"u}o{\u{g}}lu, B. C.},
    title = "{Long-lived quasinormal modes in the Euler-Heisenberg electrodynamics}",
    eprint = "2508.13361",
    archivePrefix = "arXiv",
    primaryClass = "gr-qc",
    doi = "10.1016/j.physletb.2025.140026",
    journal = "Phys. Lett. B",
    volume = "871",
    pages = "140026",
    year = "2025"
}

@article{Lutfuoglu:2025qkt,
    author = {L{\"u}tf{\"u}o{\u{g}}lu, B. C.},
    title = "{Long-lived quasinormal modes and echoes in the Einstein{\textendash}Gauss{\textendash}Bonnet{\textendash}Proca theory}",
    eprint = "2508.19194",
    archivePrefix = "arXiv",
    primaryClass = "gr-qc",
    doi = "10.1140/epjc/s10052-025-14839-x",
    journal = "Eur. Phys. J. C",
    volume = "85",
    number = "9",
    pages = "1076",
    year = "2025"
}

@article{Lutfuoglu:2025hjy,
    author = {L{\"u}tf{\"u}o{\u{g}}lu, B. C.},
    title = "{Long-lived quasinormal modes and gray-body factors of black holes and wormholes in dark matter inspired Weyl gravity}",
    eprint = "2503.16087",
    archivePrefix = "arXiv",
    primaryClass = "gr-qc",
    doi = "10.1140/epjc/s10052-025-14210-0",
    journal = "Eur. Phys. J. C",
    volume = "85",
    number = "5",
    pages = "486",
    year = "2025"
}

@article{Konoplya:2021ube,
    author = "Konoplya, R. A.",
    title = "{Black holes in galactic centers: Quasinormal ringing, grey-body factors and Unruh temperature}",
    eprint = "2109.01640",
    archivePrefix = "arXiv",
    primaryClass = "gr-qc",
    doi = "10.1016/j.physletb.2021.136734",
    journal = "Phys. Lett. B",
    volume = "823",
    pages = "136734",
    year = "2021"
}

@article{Konoplya:2019xmn,
    author = "Konoplya, R. A.",
    title = "{Quantum corrected black holes: quasinormal modes, scattering, shadows}",
    eprint = "1912.10582",
    archivePrefix = "arXiv",
    primaryClass = "gr-qc",
    doi = "10.1016/j.physletb.2020.135363",
    journal = "Phys. Lett. B",
    volume = "804",
    pages = "135363",
    year = "2020"
}

@article{Konoplya:2023moy,
    author = "Konoplya, R. A. and Zhidenko, A.",
    title = "{Analytic expressions for quasinormal modes and grey-body factors in the eikonal limit and beyond}",
    eprint = "2309.02560",
    archivePrefix = "arXiv",
    primaryClass = "gr-qc",
    doi = "10.1088/1361-6382/ad0a52",
    journal = "Class. Quant. Grav.",
    volume = "40",
    number = "24",
    pages = "245005",
    year = "2023"
}

@article{Konoplya:2024lir,
    author = "Konoplya, R. A. and Zhidenko, A.",
    title = "{Correspondence between grey-body factors and quasinormal modes}",
    eprint = "2406.11694",
    archivePrefix = "arXiv",
    primaryClass = "gr-qc",
    doi = "10.1088/1475-7516/2024/09/068",
    journal = "JCAP",
    volume = "09",
    pages = "068",
    year = "2024"
}

@article{Konoplya:2004wg,
    author = "Konoplya, R. A. and Zhidenko, A. V.",
    title = "{Decay of massive scalar field in a Schwarzschild background}",
    eprint = "gr-qc/0411059",
    archivePrefix = "arXiv",
    doi = "10.1016/j.physletb.2005.01.078",
    journal = "Phys. Lett. B",
    volume = "609",
    pages = "377--384",
    year = "2005"
}

@article{Aragon:2020teq,
    author = "Arag{\'o}n, Almendra and B{\'e}car, Ram{\'o}n and Gonz{\'a}lez, P. A. and V{\'a}squez, Yerko",
    title = "{Massive Dirac quasinormal modes in Schwarzschild{\textendash}de Sitter black holes: Anomalous decay rate and fine structure}",
    eprint = "2009.09436",
    archivePrefix = "arXiv",
    primaryClass = "gr-qc",
    doi = "10.1103/PhysRevD.103.064006",
    journal = "Phys. Rev. D",
    volume = "103",
    number = "6",
    pages = "064006",
    year = "2021"
}

@article{Ponglertsakul:2020ufm,
    author = "Ponglertsakul, Supakchai and Gwak, Bogeun",
    title = "{Massive scalar perturbations on Myers-Perry{\textendash}de Sitter black holes with a single rotation}",
    eprint = "2007.16108",
    archivePrefix = "arXiv",
    primaryClass = "gr-qc",
    doi = "10.1140/epjc/s10052-020-08616-1",
    journal = "Eur. Phys. J. C",
    volume = "80",
    number = "11",
    pages = "1023",
    year = "2020"
}

@article{Gonzalez:2022upu,
    author = "Gonz{\'a}lez, P. A. and Papantonopoulos, Eleftherios and Saavedra, Joel and V{\'a}squez, Yerko",
    title = {{Quasinormal modes for massive charged scalar fields in Reissner-Nordstr{\"o}m dS black holes: anomalous decay rate}},
    eprint = "2204.01570",
    archivePrefix = "arXiv",
    primaryClass = "gr-qc",
    doi = "10.1007/JHEP06(2022)150",
    journal = "JHEP",
    volume = "06",
    pages = "150",
    year = "2022"
}

@article{Burikham:2017gdm,
    author = "Burikham, Piyabut and Ponglertsakul, Supakchai and Tannukij, Lunchakorn",
    title = "{Charged scalar perturbations on charged black holes in de Rham-Gabadadze-Tolley massive gravity}",
    eprint = "1709.02716",
    archivePrefix = "arXiv",
    primaryClass = "gr-qc",
    doi = "10.1103/PhysRevD.96.124001",
    journal = "Phys. Rev. D",
    volume = "96",
    number = "12",
    pages = "124001",
    year = "2017"
}

@article{Seahra:2004fg,
    author = "Seahra, Sanjeev S. and Clarkson, Chris and Maartens, Roy",
    title = "{Detecting extra dimensions with gravity wave spectroscopy: the black string brane-world}",
    eprint = "gr-qc/0408032",
    archivePrefix = "arXiv",
    doi = "10.1103/PhysRevLett.94.121302",
    journal = "Phys. Rev. Lett.",
    volume = "94",
    pages = "121302",
    year = "2005"
}

@article{Churilova:2020bql,
    author = "Churilova, M. S.",
    title = "{Black holes in Einstein-aether theory: Quasinormal modes and time-domain evolution}",
    eprint = "2002.03450",
    archivePrefix = "arXiv",
    primaryClass = "gr-qc",
    doi = "10.1103/PhysRevD.102.024076",
    journal = "Phys. Rev. D",
    volume = "102",
    number = "2",
    pages = "024076",
    year = "2020"
}

@article{Zinhailo:2024jzt,
    author = "Zinhailo, Antonina F.",
    title = "{Exploring unique quasinormal modes of a massive scalar field in brane-world scenarios}",
    eprint = "2403.06867",
    archivePrefix = "arXiv",
    primaryClass = "gr-qc",
    doi = "10.1016/j.physletb.2024.138682",
    journal = "Phys. Lett. B",
    volume = "853",
    pages = "138682",
    year = "2024"
}

@article{Jing:2004zb,
    author = "Jing, Jiliang",
    title = "{Late-time evolution of charged massive Dirac fields in the Reissner-Nordstrom black-hole background}",
    eprint = "gr-qc/0408090",
    archivePrefix = "arXiv",
    doi = "10.1103/PhysRevD.72.027501",
    journal = "Phys. Rev. D",
    volume = "72",
    pages = "027501",
    year = "2005"
}

@article{Koyama:2001qw,
    author = "Koyama, Hiroko and Tomimatsu, Akira",
    title = "{Slowly decaying tails of massive scalar fields in spherically symmetric space-times}",
    eprint = "gr-qc/0112075",
    archivePrefix = "arXiv",
    doi = "10.1103/PhysRevD.65.084031",
    journal = "Phys. Rev. D",
    volume = "65",
    pages = "084031",
    year = "2002"
}

@article{Moderski:2001tk,
    author = "Moderski, Rafal and Rogatko, Marek",
    title = "{Late time evolution of a selfinteracting scalar field in the space-time of dilaton black hole}",
    eprint = "gr-qc/0105056",
    archivePrefix = "arXiv",
    doi = "10.1103/PhysRevD.64.044024",
    journal = "Phys. Rev. D",
    volume = "64",
    pages = "044024",
    year = "2001"
}

@article{Rogatko:2007zz,
    author = "Rogatko, Marek and Szyplowska, Agnieszka",
    title = "{Decay of massive scalar hair on brane black holes}",
    doi = "10.1103/PhysRevD.76.044010",
    journal = "Phys. Rev. D",
    volume = "76",
    pages = "044010",
    year = "2007"
}

@article{Koyama:2001ee,
    author = "Koyama, Hiroko and Tomimatsu, Akira",
    title = "{Asymptotic tails of massive scalar fields in Schwarzschild background}",
    eprint = "gr-qc/0103086",
    archivePrefix = "arXiv",
    doi = "10.1103/PhysRevD.64.044014",
    journal = "Phys. Rev. D",
    volume = "64",
    pages = "044014",
    year = "2001"
}

@article{Koyama:2000hj,
    author = "Koyama, Hiroko and Tomimatsu, Akira",
    title = "{Asymptotic power law tails of massive scalar fields in Reissner-Nordstrom background}",
    eprint = "gr-qc/0012022",
    archivePrefix = "arXiv",
    doi = "10.1103/PhysRevD.63.064032",
    journal = "Phys. Rev. D",
    volume = "63",
    pages = "064032",
    year = "2001"
}

@article{Gibbons:2008gg,
    author = "Gibbons, Gary W. and Rogatko, Marek and Szyplowska, Agnieszka",
    title = "{Decay of Massive Dirac Hair on a Brane-World Black Hole}",
    eprint = "0802.3259",
    archivePrefix = "arXiv",
    primaryClass = "hep-th",
    doi = "10.1103/PhysRevD.77.064024",
    journal = "Phys. Rev. D",
    volume = "77",
    pages = "064024",
    year = "2008"
}

@article{Gibbons:2008rs,
    author = "Gibbons, Gary W. and Rogatko, Marek",
    title = "{The Decay of Dirac Hair around a Dilaton Black Hole}",
    eprint = "0801.3130",
    archivePrefix = "arXiv",
    primaryClass = "hep-th",
    doi = "10.1103/PhysRevD.77.044034",
    journal = "Phys. Rev. D",
    volume = "77",
    pages = "044034",
    year = "2008"
}

@article{Konoplya:2008hj,
    author = "Konoplya, R. A.",
    title = "{Magnetic field creates strong superradiant instability}",
    eprint = "0801.0846",
    archivePrefix = "arXiv",
    primaryClass = "hep-th",
    doi = "10.1016/j.physletb.2008.11.059",
    journal = "Phys. Lett. B",
    volume = "666",
    pages = "283--287",
    year = "2008"
}

@article{Wu:2015fwa,
    author = "Wu, Chen and Xu, Renli",
    title = "{Decay of massive scalar field in a black hole background immersed in magnetic field}",
    eprint = "1507.04911",
    archivePrefix = "arXiv",
    primaryClass = "gr-qc",
    doi = "10.1140/epjc/s10052-015-3632-1",
    journal = "Eur. Phys. J. C",
    volume = "75",
    number = "8",
    pages = "391",
    year = "2015"
}

@article{Konoplya:2023fmh,
    author = "Konoplya, R. A. and Zhidenko, A.",
    title = "{Asymptotic tails of massive gravitons in light of pulsar timing array observations}",
    eprint = "2307.01110",
    archivePrefix = "arXiv",
    primaryClass = "gr-qc",
    doi = "10.1016/j.physletb.2024.138685",
    journal = "Phys. Lett. B",
    volume = "853",
    pages = "138685",
    year = "2024"
}

@article{NANOGrav:2023hvm,
    author = "Afzal, Adeela and others",
    collaboration = "NANOGrav",
    title = "{The NANOGrav 15 yr Data Set: Search for Signals from New Physics}",
    eprint = "2306.16219",
    archivePrefix = "arXiv",
    primaryClass = "astro-ph.HE",
    reportNumber = "FERMILAB-PUB-23-589-T",
    doi = "10.3847/2041-8213/acdc91",
    journal = "Astrophys. J. Lett.",
    volume = "951",
    number = "1",
    pages = "L11",
    year = "2023",
    note = "[Erratum: Astrophys.J.Lett. 971, L27 (2024), Erratum: Astrophys.J. 971, L27 (2024)]"
}

@article{Fernandes:2021qvr,
    author = "Fernandes, Tiago V. and Hilditch, David and Lemos, Jos{\'e} P. S. and Cardoso, Vitor",
    title = "{Quasinormal modes of Proca fields in a Schwarzschild-AdS spacetime}",
    eprint = "2112.03282",
    archivePrefix = "arXiv",
    primaryClass = "gr-qc",
    doi = "10.1103/PhysRevD.105.044017",
    journal = "Phys. Rev. D",
    volume = "105",
    number = "4",
    pages = "044017",
    year = "2022"
}

@article{Kokkotas:2010zd,
    author = "Kokkotas, K. D. and Konoplya, R. A. and Zhidenko, A.",
    title = "{Quasinormal modes, scattering and Hawking radiation of Kerr-Newman black holes in a magnetic field}",
    eprint = "1011.1843",
    archivePrefix = "arXiv",
    primaryClass = "gr-qc",
    doi = "10.1103/PhysRevD.83.024031",
    journal = "Phys. Rev. D",
    volume = "83",
    pages = "024031",
    year = "2011"
}

@article{Konoplya:2017lhs,
    author = "Konoplya, R. A. and Zhidenko, A.",
    title = "{The portrait of eikonal instability in Lovelock theories}",
    eprint = "1705.01656",
    archivePrefix = "arXiv",
    primaryClass = "hep-th",
    doi = "10.1088/1475-7516/2017/05/050",
    journal = "JCAP",
    volume = "05",
    pages = "050",
    year = "2017"
}

@article{Niedermaier:2006wt,
    author = "Niedermaier, Max and Reuter, Martin",
    title = "{The Asymptotic Safety Scenario in Quantum Gravity}",
    doi = "10.12942/lrr-2006-5",
    journal = "Living Rev. Rel.",
    volume = "9",
    pages = "5--173",
    year = "2006"
}

@article{Malik:2025erb,
    author = "Malik, Zainab",
    title = "{Grey-Body Factors for Scalar and Dirac Fields in the Euler-Heisenberg Electrodynamics}",
    eprint = "2509.15995",
    archivePrefix = "arXiv",
    primaryClass = "gr-qc",
    doi = "10.53941/ijgtp.2025.100006",
    journal = "Int. J. Grav. Theor. Phys.",
    volume = "1",
    pages = "6",
    year = "2025"
}

@article{Malik:2024wvs,
    author = "Malik, Zainab",
    title = "{Quasinormal modes and grey-body factors of Morris-Thorne wormholes}",
    eprint = "2412.13385",
    archivePrefix = "arXiv",
    primaryClass = "gr-qc",
    month = "12",
    year = "2024"
}

@article{Stuchlik:2025ezz,
    author = "Stuchl{\'\i}k, Z. and Zhidenko, A.",
    title = "{Overtones behavior of higher dimensional black holes in the Einstein-Gauss-Bonnet gravity}",
    eprint = "2503.06775",
    archivePrefix = "arXiv",
    primaryClass = "gr-qc",
    doi = "10.1103/qv83-1jw3",
    journal = "Phys. Rev. D",
    volume = "112",
    number = "2",
    pages = "024064",
    year = "2025"
}

@article{Stuchlik:2025mjj,
    author = "Stuchl{\'\i}k, Z. and Zhidenko, A.",
    title = {{Non-oscillatory gravitational quasinormal modes of Reissner-Nordstr{\"o}m-de Sitter spacetime}},
    eprint = "2506.09829",
    archivePrefix = "arXiv",
    primaryClass = "gr-qc",
    month = "6",
    year = "2025"
}

@article{Skvortsova:2025cah,
    author = "Skvortsova, Milena",
    title = "{Arbitrarily long-lived quasinormal modes of proper-time flow black holes}",
    eprint = "2509.18061",
    archivePrefix = "arXiv",
    primaryClass = "gr-qc",
    month = "9",
    year = "2025"
}

@article{Bolokhov:2025fto,
    author = "Bolokhov, S. V.",
    title = "{Quasinormal ringing of a regular black hole sourced by the Dehnen-type distribution of matter}",
    eprint = "2511.12859",
    archivePrefix = "arXiv",
    primaryClass = "gr-qc",
    doi = "10.1016/j.aop.2026.170416",
    journal = "Annals Phys.",
    volume = "488",
    pages = "170416",
    year = "2026"
}

@article{Bolokhov:2026eqf,
    author = "Bolokhov, S. V.",
    title = "{Quasinormal Modes and Grey-Body Factors of Scalar, Electromagnetic and Dirac Fields Around Einasto-Supported Regular Black Holes}",
    eprint = "2603.22310",
    archivePrefix = "arXiv",
    primaryClass = "gr-qc",
    month = "3",
    year = "2026"
}

@article{Burko:2004jn,
    author = "Burko, Lior M. and Khanna, Gaurav",
    title = "{Universality of massive scalar field late time tails in black hole space-times}",
    eprint = "gr-qc/0403018",
    archivePrefix = "arXiv",
    doi = "10.1103/PhysRevD.70.044018",
    journal = "Phys. Rev. D",
    volume = "70",
    pages = "044018",
    year = "2004"
}

@article{Dubinsky:2024mwd,
    author = "Dubinsky, Alexey",
    title = "{Quantum Gravitational Corrections to the Schwarzschild Spacetime and Quasinormal Frequencies}",
    eprint = "2405.13552",
    archivePrefix = "arXiv",
    primaryClass = "gr-qc",
    doi = "10.1007/s10773-025-06053-y",
    journal = "Int. J. Theor. Phys.",
    volume = "64",
    number = "8",
    pages = "203",
    year = "2025"
}

@article{Zinhailo:2018ska,
    author = "Zinhailo, A. F.",
    title = "{Quasinormal modes of the four-dimensional black hole in Einstein{\textendash}Weyl gravity}",
    eprint = "1809.03913",
    archivePrefix = "arXiv",
    primaryClass = "gr-qc",
    doi = "10.1140/epjc/s10052-018-6467-8",
    journal = "Eur. Phys. J. C",
    volume = "78",
    number = "12",
    pages = "992",
    year = "2018"
}

@article{Percival:2020skc,
    author = "Percival, Jake and Dolan, Sam R.",
    title = "{Quasinormal modes of massive vector fields on the Kerr spacetime}",
    eprint = "2008.10621",
    archivePrefix = "arXiv",
    primaryClass = "gr-qc",
    doi = "10.1103/PhysRevD.102.104055",
    journal = "Phys. Rev. D",
    volume = "102",
    number = "10",
    pages = "104055",
    year = "2020"
}

@article{Zhang:2018jgj,
    author = "Zhang, Ming and Jiang, Jie and Zhong, Zhen",
    title = {{The longlived charged massive scalar field in the higher-dimensional Reissner{\textendash}Nordstr{\"o}m spacetime}},
    eprint = "1811.04183",
    archivePrefix = "arXiv",
    primaryClass = "gr-qc",
    doi = "10.1016/j.physletb.2018.10.072",
    journal = "Phys. Lett. B",
    volume = "789",
    pages = "13--18",
    year = "2019"
}

@article{Ohashi:2004wr,
    author = "Ohashi, Akira and Sakagami, Masa-aki",
    title = "{Massive quasi-normal mode}",
    eprint = "gr-qc/0407009",
    archivePrefix = "arXiv",
    doi = "10.1088/0264-9381/21/16/010",
    journal = "Class. Quant. Grav.",
    volume = "21",
    pages = "3973--3984",
    year = "2004"
}

@article{Zhidenko:2006rs,
    author = "Zhidenko, Alexander",
    title = "{Massive scalar field quasi-normal modes of higher dimensional black holes}",
    eprint = "gr-qc/0607133",
    archivePrefix = "arXiv",
    doi = "10.1103/PhysRevD.74.064017",
    journal = "Phys. Rev. D",
    volume = "74",
    pages = "064017",
    year = "2006"
}

@article{Konoplya:2017tvu,
    author = "Konoplya, Roman A. and Zhidenko, Alexander",
    title = "{Quasinormal modes of massive fermions in Kerr spacetime: Long-lived modes and the fine structure}",
    eprint = "1712.06667",
    archivePrefix = "arXiv",
    primaryClass = "gr-qc",
    doi = "10.1103/PhysRevD.97.084034",
    journal = "Phys. Rev. D",
    volume = "97",
    number = "8",
    pages = "084034",
    year = "2018"
}

@article{Konoplya:2018qov,
    author = "Konoplya, R. A. and Stuchl{\'\i}k, Z. and Zhidenko, A.",
    title = {{Massive nonminimally coupled scalar field in Reissner-Nordstr{\"o}m spacetime: Long-lived quasinormal modes and instability}},
    eprint = "1808.03346",
    archivePrefix = "arXiv",
    primaryClass = "gr-qc",
    doi = "10.1103/PhysRevD.98.104033",
    journal = "Phys. Rev. D",
    volume = "98",
    number = "10",
    pages = "104033",
    year = "2018"
}

@article{Konoplya:2013rxa,
    author = "Konoplya, R. A. and Zhidenko, A.",
    title = "{Massive charged scalar field in the Kerr-Newman background I: quasinormal modes, late-time tails and stability}",
    eprint = "1307.1812",
    archivePrefix = "arXiv",
    primaryClass = "gr-qc",
    doi = "10.1103/PhysRevD.88.024054",
    journal = "Phys. Rev. D",
    volume = "88",
    pages = "024054",
    year = "2013"
}

@article{Malik:2023bxc,
    author = "Malik, Zainab",
    title = "{Quasinormal Modes of the Bumblebee Black Holes with a Global Monopole}",
    eprint = "2308.10412",
    archivePrefix = "arXiv",
    primaryClass = "gr-qc",
    doi = "10.1007/s10773-024-05737-1",
    journal = "Int. J. Theor. Phys.",
    volume = "63",
    number = "8",
    pages = "199",
    year = "2024"
}

@article{Bolokhov:2024otn,
    author = "Bolokhov, S. V. and Skvortsova, Milena",
    title = "{Correspondence between quasinormal modes and grey-body factors of spherically symmetric traversable wormholes}",
    eprint = "2412.11166",
    archivePrefix = "arXiv",
    primaryClass = "gr-qc",
    doi = "10.1088/1475-7516/2025/04/025",
    journal = "JCAP",
    volume = "04",
    pages = "025",
    year = "2025"
}

@article{Malik:2024nhy,
    author = "Malik, Zainab",
    title = "{Perturbations and quasinormal modes of the Dirac field in Effective Quantum Gravity}",
    eprint = "2409.01561",
    archivePrefix = "arXiv",
    primaryClass = "gr-qc",
    doi = "10.1016/j.aop.2025.170046",
    journal = "Annals Phys.",
    volume = "479",
    pages = "170046",
    year = "2025"
}

@article{Matyjasek:2017psv,
    author = "Matyjasek, Jerzy and Opala, Micha{\l}",
    title = "{Quasinormal modes of black holes. The improved semianalytic approach}",
    eprint = "1704.00361",
    archivePrefix = "arXiv",
    primaryClass = "gr-qc",
    doi = "10.1103/PhysRevD.96.024011",
    journal = "Phys. Rev. D",
    volume = "96",
    number = "2",
    pages = "024011",
    year = "2017"
}

@article{Page:1976df,
    author = "Page, Don N.",
    title = "{Particle Emission Rates from a Black Hole: Massless Particles from an Uncharged, Nonrotating Hole}",
    doi = "10.1103/PhysRevD.13.198",
    journal = "Phys. Rev. D",
    volume = "13",
    pages = "198--206",
    year = "1976"
}

@article{Page:1976ki,
    author = "Page, Don N.",
    title = "{Particle Emission Rates from a Black Hole. 2. Massless Particles from a Rotating Hole}",
    doi = "10.1103/PhysRevD.14.3260",
    journal = "Phys. Rev. D",
    volume = "14",
    pages = "3260--3273",
    year = "1976"
}

\end{document}